\documentclass[12pt]{iopart}
\usepackage[utf8]{inputenc} 
\usepackage[T1]{fontenc}
\expandafter\let\csname equation*\endcsname\relax
\expandafter\let\csname endequation*\endcsname\relax
\usepackage{amsmath}
\usepackage{hyperref} 
\usepackage{cleveref} 
\hypersetup{colorlinks = true, 
	    linkcolor = blue, 
	    urlcolor = blue,
           citecolor = blue} 
\usepackage{url}            
\usepackage{booktabs} 
\usepackage{amsfonts}       
\usepackage{nicefrac}      
\usepackage{microtype}           
\usepackage[pdftex]{graphicx}       
\usepackage{subfigure}
\usepackage{color}
\usepackage{float} 
\usepackage{amsthm}
\usepackage{soul, color, xcolor}

\graphicspath{{figs/}}

\usepackage{iopams}
\usepackage{soul}
\begin{document}

\title[]{Beyond the Dynamical Lie Algebra: Jordan-Lie Structure and a Classically Computable Witness for Variational Quantum Circuits}

\author{Yun Shang$^{1,2,4,\dag}$, Tianen Chen$^{1,3,4}$}
\address{$^1$ Institute of Mathematics, Academy of Mathematics and Systems Science, Chinese Academy of Sciences, Beijing, 100190, China}
\address{$^2$ NCMIS, MDIS, Academy of Mathematics and Systems Science, Chinese Academy of Sciences, Beijing, 100190, China}
\address{$^3$ School of Mathematical Sciences, University of Chinese Academy of Sciences, Beijing, 100190, China}
\ead{shangyun@amss.ac.cn}

\begin{abstract}
The expressive power of a variational quantum circuit is usually characterized by its dynamical Lie algebra (DLA). We show that introducing control degrees of freedom gives rise to a richer algebraic structure. Using the Hybrid Quantum Walk (HQW) model, a variational ansatz recently proposed, we prove that the algebraic closure takes the form $\mathfrak{su}(2)\otimes\mathcal{L}_Q \oplus I_c\otimes\mathcal{K}_Q$ and strictly contains the DLA of any single-path circuit. The extra directions originate from Jordan products $i\{H_c, H_b\}$ that lie outside the Lie closure and are inaccessible to single-path circuits regardless of depth. Pontryagin's minimum principle independently confirms that the optimal coin operator is generally not the Pauli-X gate, the choice that collapses HQW to Quantum Approximate Optimization Algorithm (QAOA). We introduce the Jordan product negativity $|\mathcal{N}_{\min}|$, a classically computable witness: when $|\mathcal{N}_{\min}|>0$, HQW has strictly greater expressive capacity, and its magnitude predicts the strength of this advantage. Across 90 Max-Cut instances ($n=7$ to $9$), HQW achieves a $1.60\times$ larger $d_{\rm expl}$ than QAOA in $97.8\%$ of cases, independent of graph density. In more than 100 direct benchmarks, HQW outperforms QAOA on $98\%$ of medium-density graphs and all near-complete and complete graphs. These results establish a design principle: variational quantum ans\"atze with controlled evolution paths should activate the full Jordan-Lie algebraic structure of generator Hamiltonians. Moreover, the $|\mathcal{N}_{\min}|$ witness provides a classically computable tool for identifying instances where such activation is warranted.

\end{abstract}

\maketitle

\section{Introduction}

Variational quantum algorithms (VQAs) \cite{1,64} have become a leading paradigm for near-term quantum computation \cite{63}. The expressive power of a VQA circuit is determined by its dynamical Lie algebra (DLA), the real Lie closure of its generator Hamiltonians \cite{40,35}. This Lie-algebraic framework underpins our understanding of circuit expressivity, trainability, and the onset of barren plateaus \cite{larocca2022dla,41,61,54}. Yet it captures only part of the algebraic structure that the generator Hamiltonians possess. The Hamiltonians also generate a Jordan-algebraic structure through anticommutators $i\{H_c, H_b\}$ alongside commutators $[iH_c, iH_b]$, and these Jordan product directions lie beyond the Lie closure. No existing variational ansatz systematically accesses this structure.

Recently the Hybrid Quantum Walk (HQW) \cite{31} was proposed, a model in which a coin qubit superposes multiple Hamiltonian-driven evolution paths within each circuit layer. Quantum walks support universal quantum computation \cite{27} and have been extensively studied \cite{29}. Architectures based on quantum walks have been shown to outperform QAOA in specific settings \cite{15}, but the algebraic mechanism has not been identified. We find that HQW's algebraic closure is not a standard Lie algebra but a Jordan-Lie algebra \cite{36} that strictly contains the Lie algebra of any single-path ans\"atze. Commutators in the coin space generate Jordan products $i\{H_c, H_b\}$ in the position space. This is a direct consequence of the coin-position separation and is not visible when the circuit is described as a single unitary on a monolithic register. These Jordan product directions lie outside the Lie closure. No amount of additional layers in a single-path circuit can reach them, and symmetry constraints \cite{9} impose further barriers that depth alone cannot overcome. The difference is algebraic, not a matter of depth or parameter count. In particular, QAOA \cite{2,60,37} and its multi-angle variants \cite{6} correspond to the special case where every coin is fixed to the Pauli-X gate, confining them to the Lie-algebraic component of the HQW algebra. Pontryagin's minimum principle (PMP) \cite{11}, previously applied to optimize variational quantum algorithms \cite{12} and to design optimal schedules for quantum annealing and QAOA \cite{14}, provides complementary support from optimal control theory: the optimal coin operator is generally not the Pauli-X gate, confirming that QAOA is suboptimal even within the single-path subfamily of HQW.

We make this structure rigorous through Theorem~1, which establishes that the HQW DLA takes the form $\mathfrak{g}_{\rm H} = \mathfrak{su}(2)\otimes\mathcal{L}_{\rm Q} \oplus I_c\otimes\mathcal{K}_{\rm Q}$, where $\mathcal{L}_{\rm Q}$ is the Jordan-Lie algebra generated by $\{iH_c, iH_b, iI_p\}$ and $\mathcal{K}_{\rm Q}$ is its derived Lie subalgebra. We test this prediction on over 100 problem instances. We measure the effective exploration dimension $d_{\rm expl}$, the participation ratio of the tangent-space singular values at the optimized parameters, across 90 Max-Cut instances ($n=7$--$9$, densities $0.06$--$1.0$). HQW achieves a $1.60\times$ larger $d_{\rm expl}$ than QAOA in $97.8\%$ of cases, with mean ratio $1.60\times$, independent of graph density (Pearson $r = 0.058$, $p = 0.59$). On $100+$ direct benchmarks spanning Max-Cut and Maximum Independent Set, HQW outperforms QAOA on $98\%$ of medium-density graphs and all near-complete and complete graphs. Component analysis confirms that the advantage originates from coin-controlled path superposition and cannot be replicated by reordering QAOA parameters. The algebraic advantage predicted by Theorem~1 translates consistently into measured performance gains.

These results establish a design principle that extends beyond the HQW model. The Jordan-Lie algebraic structure is a general property of any pair of non-commuting Hamiltonians. The $|\mathcal{N}_{\min}|$ witness, defined as the absolute minimum eigenvalue of the symmetrized product $H_c\circ H_b = \frac{1}{2}(H_c H_b + H_b H_c)$, provides a computationally cheap pre-screening tool, computable classically before any quantum resources are deployed, to identify instances where control-enhanced circuits are warranted.

The remainder of this paper is organized as follows. Section~\ref{Pre.} reviews the necessary preliminaries. Sections~\ref{special case}--\ref{Optimal coin} establish the theoretical foundation: we first show that QAOA is a special case of HQW, then use Pontryagin’s minimum principle to derive the optimal coin operator and prove it is generally not the Pauli-X gate. Section~\ref{DLA} develops the algebraic framework, proving via Theorem~1 that the HQW algebra is strictly larger than QAOA’s and establishing the Jordan product negativity $|\mathcal{N}_{\min}|$ as an algebraic witness. Section~\ref{num.exp.} presents comprehensive numerical experiments, including a 90-graph $d_{\rm expl}$ benchmark, $100+$ graph performance comparisons, and component analyses, all demonstrating that the algebraic advantage consistently translates into superior optimization performance. Section~\ref{con.} concludes with a summary and outlook.

\section{Preliminary}
\label{Pre.}

\subsection{Quantum approximate optimization algorithm}

\begin{figure}[htbp]
  \centering
  \includegraphics[width=0.6\textwidth]{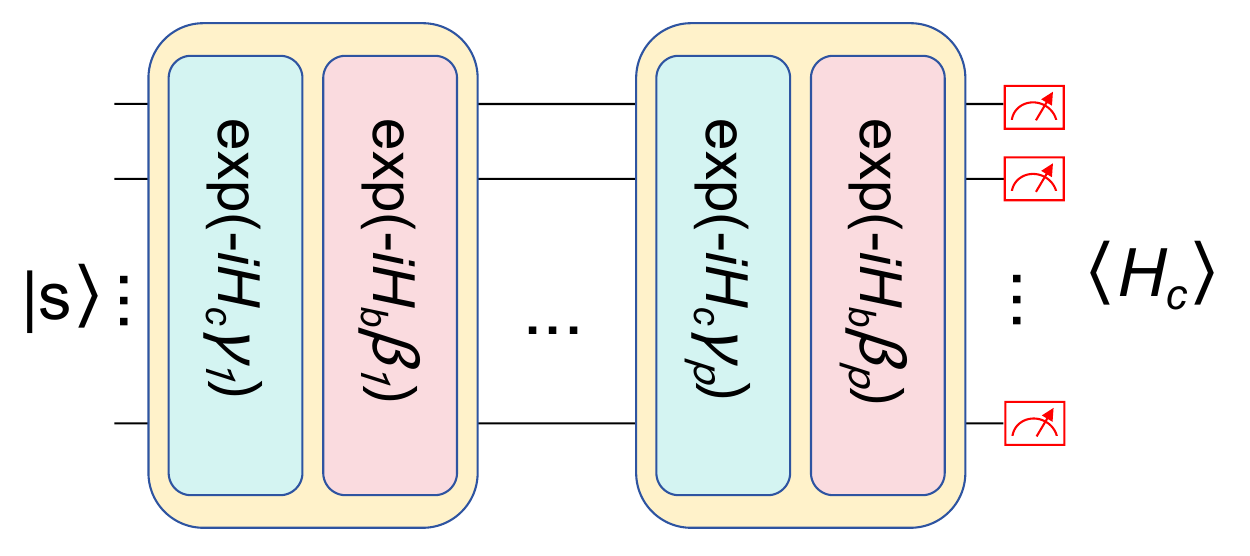}\\
  \caption{Schematic of a $p$-layer QAOA. $|s\rangle=|+\rangle^{\otimes n}$ is the initial state. $\langle H_c\rangle=\langle \psi(\boldsymbol{\beta}, \boldsymbol{\gamma}) | H_c | \psi(\boldsymbol{\beta}, \boldsymbol{\gamma}) \rangle$ is the objective function. Each yellow area corresponds to a layer of QAOA which alternately applies $e^{-iH_c\gamma_i}$ and $e^{-iH_b\beta_i}$.}
  \label{fig:5}

\end{figure}

The QAOA \cite{2} hybrid workflow, analyzed comprehensively in \cite{37}, involves preparing an ansatz state through alternating problem and mixer Hamiltonian evolutions, and using a classical optimizer to adjust the variational parameters by minimizing the expected energy, thus converging to an approximate solution. Derived as a discretization of the quantum adiabatic algorithm \cite{2,43}, QAOA leverages a parameterized quantum circuit combined with a classical optimizer, forming a closed-loop feedback system. First, QAOA proceeds by mapping the combinatorial optimization problem to a problem Hamiltonian $H_c$, whose ground state corresponds to the optimal solution. It starts with the initial state $|s\rangle=|+\rangle^{\otimes n}$ for $n$-qubit systems, the highest energy state of the mixer Hamiltonian $H_b=\sum_{i=1}^nX_i$. A parameterized quantum circuit constructs the ans\"atze state
\begin{align}
|\psi(\boldsymbol{\beta}, \boldsymbol{\gamma})\rangle = U_b(\beta_p) U_c(\gamma_p) \cdots U_b(\beta_1) U_c(\gamma_1) |s\rangle,
\end{align}
where $U_c(\gamma) = e^{-i\gamma H_c}$ and $U_b(\beta) = e^{-i\beta H_b}$ denote the problem and mixer evolution operators for $p$ layers. The classical parameters
$(\boldsymbol{\beta}, \boldsymbol{\gamma})=(\beta_1, \dots, \beta_p, \gamma_1, \dots, \gamma_p)$ are optimized by minimizing the expected cost $\langle \psi(\boldsymbol{\beta}, \boldsymbol{\gamma}) | H_c | \psi(\boldsymbol{\beta}, \boldsymbol{\gamma}) \rangle$ using methods like gradient descent. The quantum circuit of QAOA is shown as in FIG.~\ref{fig:5}. Finally, the optimized state is measured in the computational basis, and the bitstring with the highest frequency or lowest cost is identified as the approximate optimal solution.

\subsection{Hybrid quantum walk}

\begin{figure}[htbp]
  \centering
  \includegraphics[width=0.8\textwidth]{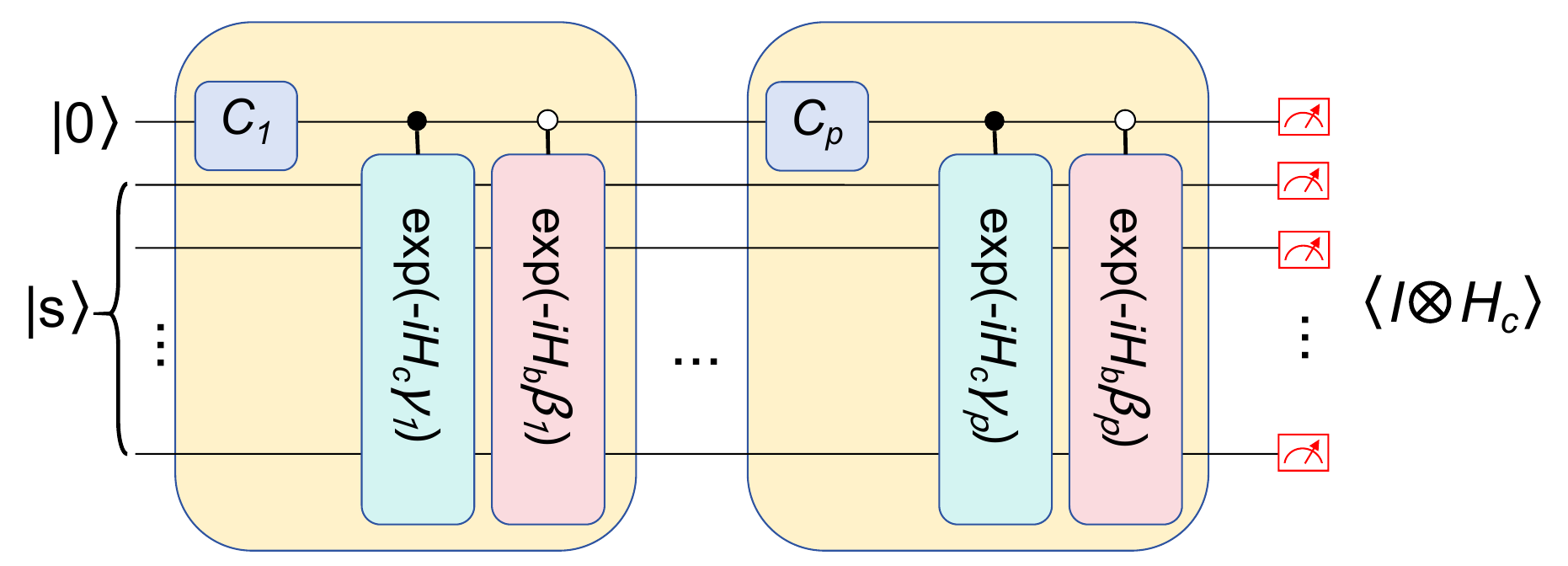}\\
  \caption{Schematic of a $p$-step HQW. $|0\rangle\otimes|s\rangle=|0\rangle\otimes|+\rangle^{\otimes n}$ is the initial state with the first qubit being the coin state and other $n$ qubits being the position space. $\langle I\otimes H_c\rangle=\langle \psi_f| H_c | \psi_f \rangle$ is the objective function. Each yellow area corresponds to a step of HQW which alternately applies the coin operator $C_i$ and the Hamiltonian evolution $e^{-i|1\rangle\langle1|\otimes H_c\gamma_i}e^{-i|0\rangle\langle0|\otimes H_b\beta_i}$.}
  \label{fig:6}
\end{figure}

The HQW model \cite{31} integrates key features from both discrete and continuous quantum walks, combining coin-based control with Hamiltonian-driven evolution to enhance performance. In this model, for a graph $G = (V, E)$, the edge set $E$ is decomposed into labeled subsets $\{E_j\}_{j \in \Gamma}$ which need not be disjoint, and the system is defined on a composite Hilbert space $\mathcal{H}_c \otimes \mathcal{H}_p = \mathrm{span}\{|j\rangle : j \in \Gamma\} \otimes \mathrm{span}\{|v\rangle:v \in V\}$. The dynamics of the HQW are governed by the operator 
\begin{align}
    W(t) = S(t)(C \otimes I) = e^{-iHt}(C \otimes I),
\end{align} 
where $C$ is a unitary coin operator acting on the coin space $\mathcal{H}_c$, $I$ is the identity on the position space $\mathcal{H}_p$, and $S(t) = e^{-iHt}$ encodes the continuous-time evolution generated by the generalized Hamiltonian
\begin{align}
H=\sum_{j\in\Gamma}|j\rangle\langle j|\otimes S_j.
\end{align} 
Here, each $S_j$ corresponds to the (weighted) adjacency matrix of the subgraph $G_j = (V, E_j)$, and the Hermiticity of $H$ necessitates that each $S_j$ be Hermitian. The quantum circuit for this HQW configuration is shown in FIG.~\ref{fig:6}. 
By introducing coin degrees of freedom and controllable subgraph structures, HQW allows more flexible quantum walk design for combinatorial optimization.

\subsection{Pontryagin’s minimum principle}

Pontryagin's Minimum Principle \cite{11} provides a necessary condition for optimality in optimal control problems. Consider a dynamical system
\begin{align}
\dot{\boldsymbol x}(t) = \boldsymbol f(\boldsymbol x(t), \boldsymbol v(t)), \quad \boldsymbol x(t_0) = \boldsymbol{x_0},    
\end{align}
where $\boldsymbol x:I\rightarrow M$ is the state with $I$ being an interval in $\mathbb{R}$ and $M$ being a smooth manifold, such as the unit sphere in $\mathbb{C}^n$. The vector $\boldsymbol{v}(t) \in \mathcal{U} \subset \mathbb{R}^m$ is the control input, and $t \in [t_0, t_f]$. The cost function to be minimized is
\begin{align}
J(\boldsymbol v) = \phi(\boldsymbol x(t_f), t_f) + \int_{t_0}^{t_f}dt g(\boldsymbol x(t),\boldsymbol{\dot x}(t),  \boldsymbol v(t), t).
\end{align}
Define the control Hamiltonian as
\begin{align}
\mathcal{H}(\boldsymbol x, \boldsymbol k, \boldsymbol v, t) = g(\boldsymbol x,\boldsymbol{\dot x} ,\boldsymbol v, t) + \boldsymbol k \cdot \boldsymbol f(\boldsymbol x, \boldsymbol v),
\end{align}
where $\boldsymbol k(t)$ is the adjoint vector. Then, if $\boldsymbol v^*(t)$ and $\boldsymbol x^*(t)$ are optimal, the following conditions hold:
\begin{enumerate}
\item $\boldsymbol{\dot x}^*(t) = \nabla_{\boldsymbol k} \mathcal{H} = \boldsymbol f(\boldsymbol x^*(t), \boldsymbol v^*(t)).$
\item $\boldsymbol{\dot k}^*(t) = -\nabla_{\boldsymbol x}\mathcal{H} = -\nabla_{\boldsymbol x}g - \boldsymbol k^*(t)\cdot \nabla_{\boldsymbol x}\boldsymbol f$.
\item $\boldsymbol k^*(t_f) = \nabla_{\boldsymbol x} \phi(\boldsymbol x^*(t_f), t_f)$.
\item $\mathcal{H}(\boldsymbol x^*(t), \boldsymbol k^*(t), \boldsymbol v^*(t)) = \min_{\boldsymbol v \in \mathcal{U}} H(\boldsymbol x^*(t), \boldsymbol k^*(t), \boldsymbol v)$
for $\text{ a.e. } t \in [t_0, t_f]$.
\end{enumerate}
These equations form a two-point boundary value problem, and the PMP often reduces the optimal control problem to solving for $\boldsymbol v^*(t)$ that minimizes the control Hamiltonian.

\subsection{Dynamical Lie algebra}

For variational quantum algorithms, the DLA characterizes the expressive power and trainability of parameterized circuits.
First, we give the definition of a Lie algebra.
\\
\textbf{Definition 1} (Lie algebra)\textbf{.}
Let $\mathfrak{g}$ be a vector space over a field $\mathbb{F}$ equipped with a bilinear map
\begin{align}
[\cdot, \cdot]: \mathfrak{g} \times \mathfrak{g} \to \mathfrak{g}, 
\end{align}
called the Lie bracket. Then $(\mathfrak{g}, [\cdot, \cdot])$ is called a Lie algebra if it satisfies the following axioms:
\begin{enumerate}
    \item[(1).] $[X, Y] = -[Y, X], \forall X, Y \in \mathfrak{g}$.
    \item[(2).] $[X, [Y, Z]] + [Y, [Z, X]] + [Z, [X, Y]] = 0, \forall  X, Y, Z \in \mathfrak{g}$.
\end{enumerate}
For an $n$-qubit VQC corresponding to a parameterized unitary operator $U(\boldsymbol{\theta})=\prod_{l=1}^L(\prod_{j=1}^Me^{i\theta_{l,j}H_j})\in \text{U}(2^n)$ with variational parameters $\boldsymbol{\theta}$ and a set of Hamiltonians $\mathcal{G}=\{H_1, H_2, \dots, H_M\}$, the DLA $\mathfrak{g}$ is defined as the Lie algebra generated by the elements of $\{iH_1, iH_2, \dots, iH_M\}$ over field $\mathbb{R}$:
\begin{align}
\mathfrak{g}&=\langle \{iH_1, iH_2, \dots, iH_M\}\rangle_{\text{Lie},\mathbb{R}}.
\end{align}
In particular, $\text{span}\{iH_k, [iH_j, iH_k], [iH_j, [iH_k, iH_l]], \dots \}$ is $\mathfrak{g}$, where $[A,B] = AB - BA$ denotes the commutator. The DLA $\mathfrak{g}$ determines the set of unitary operations $U(\boldsymbol{\theta})$ reachable. The dimension and structure of $\mathfrak{g}$ directly influence the circuit's capacity to represent diverse quantum states. If $\mathfrak{g}$ is full-dimensional, i.e., $\dim(\mathfrak{g}) = 4^n - 1$, the VQC can generate any unitary in $\text{SU}(2^n)$ up to a global phase, implying maximal expressivity. Analyzing the DLA is therefore a systematic way to understand the capabilities and limitations of variational quantum algorithms.

\section{QAOA as a special case of HQW}
\label{special case}
The HQW model combines coin-based discrete quantum walks with Hamiltonian-driven continuous quantum walks. For a graph $G=(V,E)$, the edge set is decomposed into labeled subsets $\{E_{j}:j\in\Gamma\}$. The system evolves in the composite Hilbert space $\mathcal{H}_{c}\otimes\mathcal{H}_{p} = \text{span}\{|j\rangle : j\in \Gamma\} \otimes \text{span}\{|v\rangle : v\in V\}$. The dynamics are governed by $W(t) = e^{-iHt}(C \otimes I)$, where $C$ is a unitary coin operator on $\mathcal{H}_{c}$, and $H = \sum_{j\in \Gamma} |j\rangle \langle j| \otimes S_{j}$ with each $S_{j}$ being the (weighted) adjacency matrix of the subgraph $G_{j}=(V,E_{j})$.

Treating QAOA as a special case of HQW shows the greater generality of the HQW framework, and is also needed before optimal control theory can be applied to find the best ansatz within HQW. We can insert two Hamiltonians $H_{c}$ and $H_{b}$ of QAOA to the HQW as $|1\rangle\langle1|\otimes H_{c}$ and $|0\rangle\langle0|\otimes H_{b}$. The HQW operator is
\begin{align}
W(t) &=e^{-i(|1\rangle\langle1|\otimes H_{c}+|0\rangle\langle0|\otimes H_{b})t}(C\otimes I)\nonumber \\
&=e^{-i|1\rangle\langle1|\otimes H_{c}t}e^{-i|0\rangle\langle0|\otimes H_{b}t}(C\otimes I).
\end{align}
Through the decomposition of the aforementioned operator $W(t)$, HQW can be regarded as the evolution of both Hamiltonians over a time $t$. Similar to the scenario in QAOA where the two Hamiltonians have different evolution times, we can generalize the HQW operator as follows:
\begin{equation}
W(\gamma,\beta)=e^{-i|1\rangle\langle1|\otimes H_{c}\gamma}e^{-i|0\rangle\langle0|\otimes H_{b}\beta}(C\otimes I).
\end{equation}
This generalization is justified by scaling the two Hamiltonians with parameters $\omega_{c}(\tau)=\gamma/t$ and $\omega_{b}(\tau)=\beta/t$ for an HQW of total evolution time $\tau$. Thus we can obtain
\begin{align}
W(\gamma,\beta)&=e^{-i(|1\rangle\langle1|\otimes \omega_c(\tau)H_c+|0\rangle\langle0|\otimes\omega_b(\tau)H_b)t}(C\otimes I)\nonumber\\
&=e^{-i|1\rangle\langle1|\otimes H_c\gamma}e^{-i|0\rangle\langle0|\otimes H_b\beta}(C\otimes I).
\end{align}
Let the number of HQW steps be $p$. For the $k$-th step, the evolution times are $\gamma_k$ and $\beta_k$ and the coin operator is $C_k$. The initial state is $|0\rangle\otimes |s\rangle$, then
\begin{align}
|\psi_f\rangle&=W(\gamma_{p}, \beta_{p})\cdots W(\gamma_1, \beta_1)|0\rangle\otimes|s\rangle\nonumber\\
&=\sum_{\boldsymbol{v}\in \mathbb{F}_2^p}\alpha_{\boldsymbol{v}}|v_p\rangle\otimes\prod_{k=1}^p (v_kU_c(\gamma_k)+(1-v_k)U_b(\beta_k))|s\rangle,
\end{align}
where $\alpha_{\boldsymbol{v}}$ is $\langle v_p|C_p|v_{p-1}\rangle\cdots\langle v_1|C_1|0\rangle$. The coefficient $\alpha_{\boldsymbol{v}}$ is set by the sequence of coin operators $C_{k}$, which gives coherent control over the superposition. The final state is therefore a superposition of different evolution paths. To approximate the energy of $H_{c}$, we use $\langle I\otimes H_{c}\rangle=\langle\psi_{f}|I\otimes H_{c}|\psi_{f}\rangle$ as the objective function.
If all coin operators are the Pauli-X gate and each $W(\gamma_k,\beta_k)$ is applied twice per step, the final state $|\psi_{f}\rangle$ reduces to exactly the QAOA ansatz:
\begin{align}
&W(\gamma_{p},\beta_{p})W(\gamma_{p},\beta_{p})\cdots W(\gamma_{1},\beta_{1})W(\gamma_{1},\beta_{1})|0\rangle\otimes |s\rangle\nonumber\\ 
&=|0\rangle\otimes e^{-iH_{b}\beta_{p}}e^{-iH_{c}\gamma_{p}}\cdots e^{-iH_{b}\beta_{1}}e^{-iH_{c}\gamma_{1}}|s\rangle \nonumber\\
&=|0\rangle\otimes U_{b}(\beta_{p})U_{c}(\gamma_{p})\cdots U_{b}(\beta_{1})U_{c}(\gamma_{1})|s\rangle.
\end{align}
QAOA is therefore a special case of HQW. Several QAOA variants, including multi-angle QAOA, can similarly be obtained within the HQW framework with appropriate coin operators. Prior work on QAOA optimality has examined only the single evolution path $U_{b}(\beta_{p})U_{c}(\gamma_{p})\cdots U_{b}(\beta_{1})U_{c}(\gamma_{1})$. In HQW, QAOA corresponds to the case where every coin operator is the Pauli-X gate. The Pauli-X coin produces a single interleaved ordering of the two Hamiltonians. It generates no superposition of distinct paths, so there is no reason to assume this special case is optimal. We now use the more general HQW framework to find the optimal coin operator.

\section{Optimal coin operators}
\label{Optimal coin}

Each coin operator $C_i$ is a $2\times 2$ unitary, which can be parameterized as a rotation
\begin{equation}
C_{i}=e^{-i(n_{x}^{(i)}X+n_{y}^{(i)}Y+n_{z}^{(i)}Z)\frac{\theta_{i}}{2}},
\end{equation}
where $(n_{x}^{(i)})^{2}+(n_{y}^{(i)})^{2}+(n_{z}^{(i)})^{2}=1$. To determine the optimal coin, we apply the PMP to the HQW dynamics. We design a time-dependent Hamiltonian that interpolates through three stages: coin rotation, evolution under $H_b$, and evolution under $H_c$, controlled by a parameter $u(t)\in\{0,\frac12,1\}$. This makes HQW a discrete analogue of quantum adiabatic evolution with an intermediate Hamiltonian \cite{43,45,46}, while QAOA corresponds to the case without the intermediate stage.

Applying PMP (see ~\ref{app:pmp} for the full derivation), the optimal coin rotation axis aligns with the instantaneous sensitivity vector in the coin space:
\begin{align}
n_{x}(t)&=\frac{\Phi_{X\otimes I}(t)}{\sqrt{\Phi_{X\otimes I}(t)^{2}+\Phi_{Y\otimes I}(t)^{2}+\Phi_{Z\otimes I}(t)^{2}}}, \\
n_{y}(t)&=\frac{\Phi_{Y\otimes I}(t)}{\sqrt{\Phi_{X\otimes I}(t)^{2}+\Phi_{Y\otimes I}(t)^{2}+\Phi_{Z\otimes I}(t)^{2}}}, \\
n_{z}(t)&=\frac{\Phi_{Z\otimes I}(t)}{\sqrt{\Phi_{X\otimes I}(t)^{2}+\Phi_{Y\otimes I}(t)^{2}+\Phi_{Z\otimes I}(t)^{2}}}.
\end{align}
The optimal coin is therefore generally not the Pauli-X gate. This shows that the path-superposition mechanism of HQW may offer advantages over QAOA's single-path evolution, a hypothesis we test numerically in Sec.~\ref{num.exp.}.

Although PMP gives the optimal coin axis in closed form, solving the full two-point boundary value problem is computationally prohibitive for the system sizes we consider. The objective function $\langle I\otimes H_c\rangle$ is smooth and differentiable, and gradients can be computed via the parameter-shift rule or automatic differentiation. Gradient-based optimizers such as Adam \cite{53} can therefore adjust the coin parameters numerically. PMP analysis yields three design guidelines that go beyond justifying a general rotation. First, QAOA's fixed Pauli-X coin is provably suboptimal: the optimal coin axis depends on the instantaneous state, which Pauli-X ignores. No reparameterization of $\beta$ and $\gamma$ can compensate for a structurally wrong coin. Second, the optimal coin axis aligns with the instantaneous sensitivity vector $(\Phi_X, \Phi_Y, \Phi_Z)$ in the coin space. This geometric picture explains what numerical optimizers are learning at each step. Third, because the sensitivity vector changes as the state evolves, the optimal coin is necessarily step-dependent. This justifies the independent per-step parameterization of HQW.

\section{Dynamical Lie Algebra and the Jordan Witness}
\label{DLA}

The difference in expressive capacity between QAOA and HQW has an algebraic origin. We compute the DLAs of both circuits, prove that HQW generates a strictly larger algebra, and show that the Jordan product negativity is a computable witness for this advantage. Numerical measurements of $d_{\rm expl}$ on 90 graph instances confirm that the algebraic gap is consistently realized.

\subsection{Dynamical Lie Algebras of QAOA and HQW}

For QAOA, the DLA is generated by the Lie closure of the problem and mixer Hamiltonians:
\begin{equation}
\mathfrak{g}_{\rm Q}=\langle iH_{c},iH_{b}\rangle_{\rm Lie}\subseteq\mathfrak{u}(N),
\end{equation}
where $N$ is the dimension of the Hilbert space.

HQW employs an enlarged system with a coin space and a position space, with evolution driven by conditionally applied Hamiltonians. We characterize its algebraic structure in Theorem~1 (proved in Appendix~\ref{app:dla}):

\textbf{Theorem 1.} The dynamical Lie algebra of HQW is $\mathfrak{g}_{\text{H}}=\mathfrak{su}(2)\otimes\mathcal{L}_{\text{Q}}\oplus I_{c}\otimes\mathcal{K}_{\text{Q}}$, where $I_{c}$ is the identity operator of the coin space, $\mathcal{L}_{\rm Q}$ is the Jordan-Lie algebra generated by $\{iH_{c},iH_{b},iI_{p}\}$ under the commutator $[\cdot,\cdot]$ and the Jordan product $i\{\cdot,\cdot\}$, and $\mathcal{K}_{\rm Q}=[\mathcal{L}_{\rm Q},\mathcal{L}_{\rm Q}]\oplus\operatorname{span}\{iH_{c},iH_{b}\}$.

The derivation is given in ~\ref{app:dla}. When the Jordan product $i\{H_{c},H_{b}\}$ lies outside $\mathfrak{g}_{\rm Q}$, the algebra $\mathcal{L}_{\rm Q}$ is strictly larger than $\mathfrak{g}_{\rm Q}$, giving
\begin{align}
 \dim(\mathfrak{g}_{\text{H}}) &= \dim(\mathfrak{su}(2)) \times \dim(\mathcal{L}_\text{Q})+ \dim(\mathcal{K}_\text{Q}) \nonumber\\
&>3\dim(\mathfrak{g}_\text{Q})+\dim(\mathfrak{g}_\text{Q})=4\dim(\mathfrak{g}_\text{Q}).
\end{align}
If the coin did not affect the position-space dynamics, then $\dim(\mathfrak{g}_{\rm H})=\dim(\mathfrak{u}(2))\times\dim(\mathfrak{g}_{\rm Q})=4\dim(\mathfrak{g}_{\rm Q})$. The strict inequality $\dim(\mathfrak{g}_{\rm H})>4\dim(\mathfrak{g}_{\rm Q})$ shows that the coin enriches the dynamical evolution by generating Jordan product elements absent from~$\mathfrak{g}_{\rm Q}$.

To illustrate this gap concretely, we numerically compute the algebra dimensions for three small Max-Cut instances: $P_3$, $P_4$, and $P_5$, 3, 4, and 5-vertex path graphs. For each, the QAOA DLA $\mathfrak{g}_Q = \langle iH_c, iH_b\rangle_{\rm Lie}$ and the Jordan-Lie algebra $\mathcal{L}_Q = \langle iH_c, iH_b, iI\rangle_{[\cdot,\cdot],\,i\{\cdot,\cdot\}}$ are computed by iterative commutator and Jordan product closure, with dimension determined via singular value decomposition of the vectorized operator basis. $K_Q = [\mathcal{L}_Q,\mathcal{L}_Q] \oplus \operatorname{span}\{iH_c,iH_b\}$ is obtained from commutators of the $\mathcal{L}_Q$ basis, and the HQW DLA dimension follows from $\dim(\mathfrak{g}_H) = 3\dim(\mathcal{L}_Q)+\dim(K_Q)$.

\begin{table}[ht]
  \centering
  
  \begin{tabular}{lcccc}
    \toprule
    & $P_3$ ($n=3$) & $P_4$ ($n=4$) & $P_5$ ($n=5$) \\
    \midrule
    $\dim(\mathfrak{g}_Q)$      & 10  & 16  & 26  \\
    $\dim(\mathcal{L}_Q)$       & 20  & 70  & 102 \\
    $\dim(K_Q)$                 & 18  & 66  & 192 \\
    \addlinespace
    $\dim(\mathfrak{g}_H)$
      \raisebox{2pt}{$= 3\dim(\mathcal{L}_Q) + \dim(K_Q)$}
                                & 78  & 276 & 498 \\
    $\dim(\mathfrak{g}_H) / \dim(\mathfrak{g}_Q)$
                                & 7.8$\times$ & 17.3$\times$ & 19.15$\times$ \\
    $\dim(\mathfrak{g}_H) / \max\dim(\mathfrak{su}(2^{n+1}))$
                                & 30.6\% & 27.0\% & 12.2\% \\
    \addlinespace
    Maximum possible
      $\dim(\mathfrak{g}_H)$
      \raisebox{2pt}{($\mathfrak{su}(2^{n+1})$)}
                                & 255 & 1023 & 4095 \\
    \bottomrule
  \end{tabular}
  \caption{DLA dimensions for three Max-Cut path-graph instances. In all cases $\dim(\mathfrak{g}_H) > 4\dim(\mathfrak{g}_Q)$. The excess is attributable to Jordan product elements $i\{H_c,H_b\} \in \mathcal{L}_Q$ lying outside $\mathfrak{g}_Q$. The ratio $\dim(\mathfrak{g}_H)/\dim(\mathfrak{g}_Q)$ decelerates from $+9.5\times$ ($n=3\to4$) to $+1.85\times$ ($n=4\to5$), while the fraction of $\mathfrak{su}(2^{n+1})$ occupied by $\mathfrak{g}_H$ falls from $30.6\%$ to $12.2\%$. Both trends match the behavior of $|\mathcal{N}_{\min}|$ for this Hamiltonian pair. The observed decay is a property of the standard Max-Cut encoding with an $X$-mixer and can be avoided through alternative encodings or mixer design (~\ref{app:exploration}).}
  \label{tab:dla_dims}
\end{table}

In all three cases $\mathcal{L}_Q$ is strictly larger than $\mathfrak{g}_Q$. The deceleration of the ratio growth mirrors $|\mathcal{N}_{\min}|$ for this Hamiltonian pair. For the standard Max-Cut encoding with an $X$-mixer, $|\mathcal{N}_{\min}|$ roughly halves with each added qubit, and the fraction of $\mathfrak{su}(2^{n+1})$ occupied by $\mathfrak{g}_H$ shrinks accordingly. This decay has a clear structural origin and is not a fundamental property of the Jordan-Lie framework. For the Max-Cut problem Hamiltonian $H_c = \sum_{(i,j)\in E} (I - Z_i Z_j)/2$ and the $X$-mixer $H_b = \sum_i X_i$, the Jordan product $H_c \circ H_b$ receives non-zero contributions from two sources. The identity component of each $H_c$ term contributes $\{I, X_i\} = 2X_i$, while the $Z_i Z_j$ component contributes $\{Z_i Z_j, X_k\}$, which vanishes when $k = i$ or $k = j$ (since $Z$ and $X$ anticommute) and equals $2Z_i Z_j X_k$ otherwise. Each edge therefore contributes $n-2$ non-trivial Pauli strings. For a path graph with $|E| = n-1$ edges, the Jordan product comprises $\mathcal{O}(n^2)$ distinct Pauli strings, each with $2^n$ non-zero entries. After normalization by the Frobenius norm, the eigenvalues of $H_c \circ H_b$ are diluted across an exponentially growing Hilbert space of dimension $2^n$, so $|\mathcal{N}_{\min}|$ decays with $n$.

This decay is specific to the Hamiltonian pair, not intrinsic to the witness. Numerical tests across alternative Hamiltonian structures confirm that $|\mathcal{N}_{\min}|$ at a fixed $n$ can vary by factors of 2--5 depending on the choice of $H_c$ and $H_b$ (~\ref{app:nmin_scaling}). The witness is therefore best understood as a diagnostic for a given problem encoding: $|\mathcal{N}_{\min}| > 0$ guarantees that Jordan-product enrichment exists, and its magnitude indicates the effective strength of the enrichment for that specific Hamiltonian pair. The decay observed for the standard Max-Cut $+$ $X$-mixer encoding reflects the two-local structure of this specific Hamiltonian pair. The tests in Appendix~\ref{app:nmin_scaling} confirm this encoding-specificity: the decay rate varies with the choice of $H_c$ and $H_b$, indicating that the decay is not a fundamental limitation of the witness.

\subsection{The Jordan Product Direction and Its Orthogonal Component}

The element that distinguishes $\mathcal{L}_{\rm Q}$ from $\mathfrak{g}_{\rm Q}$ is the Jordan product direction
\begin{equation}
    J \equiv -i\{iH_c, iH_b\} = i(H_c H_b + H_b H_c) = i\{H_c, H_b\}.
\end{equation}
For the Max-Cut problem $H_c$ and $H_b$ are real symmetric, so $J$ is purely imaginary symmetric. QAOA cannot generate arbitrary superpositions of $J$ with other algebra elements, because only the projection of $J$ onto $\mathfrak{g}_{\rm Q}$ is reachable via commutators. To see this, let $\mathcal{S}$ and $\mathcal{A}$ denote the spaces of real symmetric and real antisymmetric matrices. Since $iH_c, iH_b \in i\mathcal{S}$ and Lie brackets act as
$[i\mathcal{S}, i\mathcal{S}] \subseteq \mathcal{A}$,
$[i\mathcal{S}, \mathcal{A}] \subseteq i\mathcal{S}$,
$[\mathcal{A}, \mathcal{A}] \subseteq \mathcal{A}$,
the QAOA DLA decomposes as $\mathfrak{g}_{\rm Q} \subseteq i\mathcal{S} \oplus \mathcal{A}$. For any $X = iS + A \in \mathfrak{g}_{\rm Q}$ with $S\in\mathcal{S}, A\in\mathcal{A}$, the Hilbert--Schmidt inner product with $J$ gives
$\langle X, J\rangle_{\rm HS} = i\operatorname{Tr}(S\{H_c,H_b\})$,
because the antisymmetric part $\operatorname{Tr}(A\{H_c,H_b\})$ vanishes. Hence $J$ is not fully orthogonal to $\mathfrak{g}_{\rm Q}$, but the component
$J_{\perp} = J - P_{\mathfrak{g}_{\rm Q}}(J)$,
where $P_{\mathfrak{g}_{\rm Q}}$ is the orthogonal projector onto $\mathfrak{g}_{\rm Q}$ with respect to the Hilbert--Schmidt inner product, lies outside the Lie algebra and is inaccessible to QAOA regardless of depth.

In the enlarged space, let $\mathcal{M}_{\rm Q}'=\exp(I_c\otimes\mathfrak{g}_{\rm Q})$ and $\mathcal{M}_{\rm H}=\exp(\mathfrak{g}_{\rm H})$ denote the evolution manifolds reachable by QAOA and HQW, respectively. The Jordan product elements in $\mathcal{L}_{\rm Q}$ generate directions transversal to $\mathcal{M}_{\rm Q}'$. At any point $U\in\mathcal{M}_{\rm H}$, the tangent space decomposes as
\begin{equation}
T_{U}\mathcal{M}_{\rm H}=T_{U}\mathcal{M}_{\rm Q}'\oplus\mathcal{T}_{U},
\end{equation}
where $\mathcal{T}_{U}$ is the subspace spanned by the Jordan product directions and is orthogonal to $T_{U}\mathcal{M}_{\rm Q}'$. HQW can therefore move along directions orthogonal to QAOA's constrained manifold, accessing a strictly larger set of states. This geometric fact underlies the expressivity advantage: $\dim(\mathfrak{g}_{\rm H}) > 4\dim(\mathfrak{g}_{\rm Q})$ whenever $J_{\perp} \neq 0$, which holds for non-commuting $H_c$ and $H_b$.

\subsection{Jordan Product Negativity as an Algebraic Witness}

The Jordan product negativity \cite{57} formalizes this geometric observation as a computable criterion. It is the minimum eigenvalue of the symmetrized product:
\begin{equation}
\mathcal{N}_{\min} = \min{\lambda(H_c \circ H_b)},\qquad
H_c \circ H_b = \frac{1}{2}(H_c H_b + H_b H_c).
\label{eq:jordan_negativity}
\end{equation}
We use normalized Hamiltonians $\tilde{H}_c = H_c / \|H_c\|_F$, $\tilde{H}_b = H_b / \|H_b\|_F$ for scale invariance. For QAOA-type problems where $H_b$ has zero diagonal in the basis that diagonalizes $H_c$, one has $\operatorname{Tr}(H_c \circ H_b) = 0$, so the eigenvalues of $H_c \circ H_b$ sum to zero and $\mathcal{N}_{\min} \in [-1, 0]$. We take the absolute value $|\mathcal{N}_{\min}|$ as the relevant quantity because only the magnitude controls the condition $J_\perp \neq 0$. The sign of the minimum eigenvalue carries no algebraic significance. This also makes the witness applicable to general Hamiltonians for which the trace of $H_c \circ H_b$ may not vanish. When $|\mathcal{N}_{\min}| > 0$, the Jordan product $i\{H_c, H_b\}$ is non-vanishing. The geometric picture established above shows that the orthogonal component $J_\perp = J - P_{\mathfrak{g}_{\rm Q}}(J)$ is nonzero whenever this Jordan product content is not fully captured by the Lie closure $\mathfrak{g}_{\rm Q}$. In the Max-Cut setting with $\operatorname{Tr}(H_c\circ H_b)=0$, $|\mathcal{N}_{\min}| > 0$ is equivalent to $H_c\circ H_b \neq 0$, so the witness correctly flags algebraic enrichment. For general Hamiltonians, $|\mathcal{N}_{\min}| > 0$ remains a necessary condition for $J_\perp \neq 0$ and a practical indicator that the Jordan product carries content beyond $\mathfrak{g}_{\rm Q}$. Computing $\mathcal{N}_{\min}$ requires only a one-time diagonalization of $H_c H_b + H_b H_c$, a classical pre-screening step that does not involve the quantum circuit.

To verify that this algebraic capacity is realized in practice, we measure the \emph{effective exploration dimension} at the optimized circuit parameters. For each variational parameter $\theta_i$ at the optimum $\boldsymbol{\theta}^*$, we perturb by $+\varepsilon$ and compute the phase-aligned finite-difference tangent vector $\boldsymbol{t}_i = (|\psi(\theta_i+\varepsilon)\rangle - |\psi(\boldsymbol{\theta}^*)\rangle)/\varepsilon$. The singular values $\sigma_i$ of the matrix $Q = [\boldsymbol{t}_1, \ldots, \boldsymbol{t}_m]$ capture the locally accessible directions, and we quantify the effective dimension via the participation ratio
\begin{equation}
d_{\rm expl} = \frac{(\sum_i \sigma_i)^2}{\sum_i \sigma_i^2},
\label{eq:eff_rank}
\end{equation}
which varies continuously between $1$ and $m$. Details of the computation are provided in ~\ref{app:exploration}.

We benchmarked $d_{\rm expl}$ on 90 Max-Cut instances spanning $n \in \{7, 8, 9\}$ and densities $0.06$--$1.0$, comprising complete graphs $K_n$, near-complete graphs, i.e. $K_n$ with $1$--$2$ edges removed, and Erd\H{o}s--R\'{e}nyi graphs with $p \in \{0.3, 0.5, 0.7, 0.9\}$. Optimization used Adam with 150 steps, learning rate $0.1$, 5 random restarts, $p_{\rm QAOA}=2$ and $p_{\rm HQW}=4$.

\begin{figure}[htbp]
  \centering
  \includegraphics[width=\textwidth]{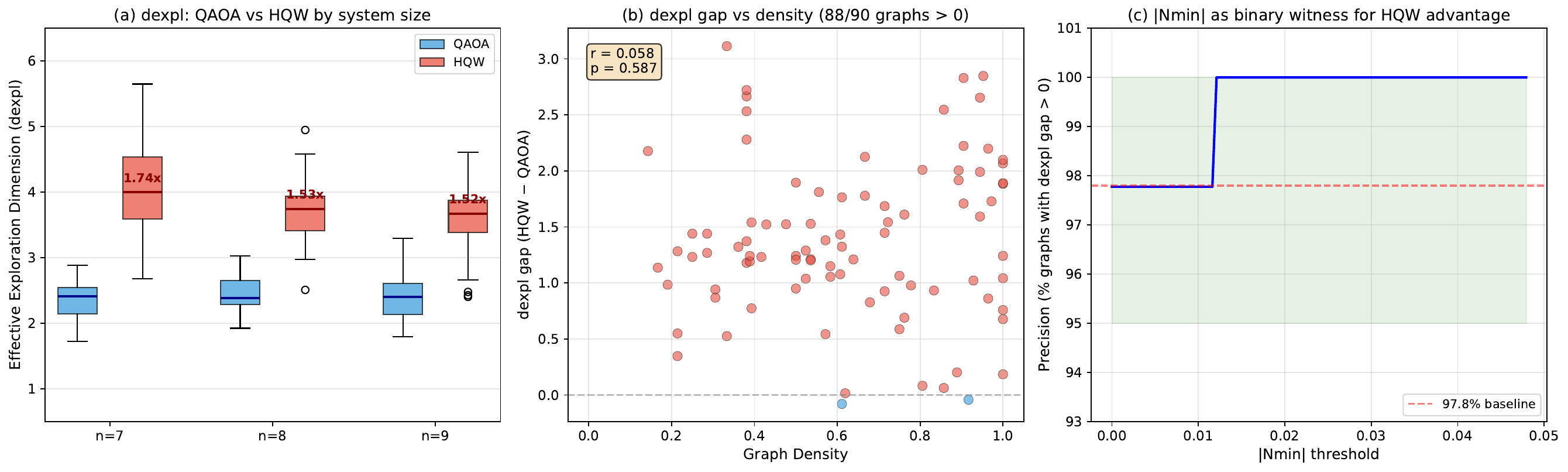}
  \caption{Effective exploration dimension $d_{\rm expl}$ (Eq.~\ref{eq:eff_rank}) across 90 Max-Cut instances with $n \in \{7, 8, 9\}$ and densities $0.06$--$1.0$. \textbf{(a)} Boxplots of $d_{\rm expl}$ for QAOA (blue) and HQW (red) by system size $n$. HQW consistently shows higher $d_{\rm expl}$, with mean ratios of $1.74\times$, $1.53\times$, and $1.52\times$ for $n=7,8,9$ respectively. \textbf{(b)} $\Delta d_{\rm expl} = d_{\rm expl}^{\rm H} - d_{\rm expl}^{\rm Q}$ versus graph density. No significant correlation is observed (Pearson $r = 0.058$, $p = 0.59$). 88/90 graphs ($97.8\%$) lie above the $\Delta d_{\rm expl} = 0$ line. \textbf{(c)} Precision of $|\mathcal{N}_{\min}| > 0$ as a binary witness for $d_{\rm expl}^{\rm H} > d_{\rm expl}^{\rm Q}$.}
  \label{fig:4}
\end{figure}

HQW has strictly greater $d_{\rm expl}$ than QAOA in 88 out of 90 graphs, with a mean ratio of $1.60\times$ (median $1.55\times$, IQR $[1.21, 1.86]$). The two counterexamples differ by less than $0.1$, within numerical noise. The $d_{\rm expl}$ gap is independent of graph density (Pearson $r = 0.058$, $p = 0.59$, Spearman $\rho = 0.036$, $p = 0.74$), confirming that the advantage is algebraic rather than density-driven. The gap varies with system size, largest at $n=7$ (mean $\Delta d_{\rm expl}=1.69$, ratio $1.74\times$), decreasing to $1.25$ ($1.53\times$) at $n=8$ and $1.16$ ($1.52\times$) at $n=9$, reflecting the exponential growth of the Hilbert space, which dilutes the fixed contribution from the Jordan product direction.

Since $d_{\rm expl} \leq m$ by construction, the higher parameter budget of HQW raises its theoretical ceiling, here $m_{\rm HQW}=12$ vs. $m_{\rm QAOA}=4$. Per-parameter efficiency, however, tells the opposite story: QAOA achieves $d_{\rm expl}^{\rm Q}/m_{\rm Q} = 0.70 \pm 0.14$, while HQW achieves $d_{\rm expl}^{\rm H}/m_{\rm H} = 0.35 \pm 0.08$. QAOA thus uses each parameter roughly twice as efficiently in terms of tangent-space dimension explored, yet HQW still wins in absolute $d_{\rm expl}$. If parameter count alone determined the outcome, we would expect a $3\times$ ratio matching the parameter ratio and $100\%$ of graphs favoring HQW, neither of which holds. The per-parameter efficiency gap is consistent with the algebraic picture: HQW's extra parameters probe a larger DLA where the signal is diluted across more dimensions. The two counterexamples, where $d_{\rm expl}^{\rm H} \leq d_{\rm expl}^{\rm Q}$ despite HQW's $3\times$ parameter advantage, further confirm that parameter count alone does not guarantee the result. We also note that perfectly equalizing parameter count, circuit depth, and Hilbert-space dimension simultaneously is structurally impossible for ans\"atze with different Lie-algebraic generators, and this asymmetry is common across the VQA comparison literature.

The Jordan product negativity $|\mathcal{N}_{\min}|$ is therefore a practical binary witness: whenever $|\mathcal{N}_{\min}| > 0$, the Jordan product is active and the geometric analysis of Sec.~\ref{DLA} predicts that HQW has strictly greater expressive capacity than QAOA. The 90-graph benchmark confirms this: $|\mathcal{N}_{\min}| > 0$ for all 90 instances, and HQW has strictly greater $d_{\rm expl}$ in $97.8\%$ of cases.

We check the negative side of the witness: $|\mathcal{N}_{\min}| = 0 \Rightarrow$ no HQW advantage, in two ways. First, we take the commuting limit $H_c = H_b = \sum_i X_i$. Here $|\mathcal{N}_{\min}| = 0$ for even $n$, since $\sum_i X_i$ has a zero eigenvalue and $|\mathcal{N}_{\min}| > 0$ for odd $n$, but commutation prevents any Jordan-product enrichment of $\mathcal{L}_{\rm Q}$ regardless of $n$. Numerical evaluation for $n=5,6,7$, where $p_{\rm QAOA}=2$ and $p_{\rm HQW}=4$, 5 restarts and 100 Adam steps, yields $d_{\rm expl}^{\rm Q} = 1.0000$ and $d_{\rm expl}^{\rm H} \in [1.0022, 1.0031]$, both reaching the exact ground-state energy $-n$. The $d_{\rm expl}$ gap thus collapses to within numerical noise regardless of $n$ parity, confirming the witness prediction. Second, for Max-Cut on non-trivial connected graphs, $|\mathcal{N}_{\min}|$ is always strictly positive because $\operatorname{Tr}(H_c \circ H_b) = 0$ forces the eigenvalues of the Jordan product to take both signs, with $|\mathcal{N}_{\min}|$ decreasing as the Hilbert space grows from $\sim\!0.042$ at $n=7$ to $\sim\!0.006$ at $n=10$. Across this range, the $d_{\rm expl}$ advantage narrows monotonically with $|\mathcal{N}_{\min}|$, consistent with the witness prediction. The witness is therefore most useful for identifying cases where $H_c$ and $H_b$ commute or nearly commute and HQW would offer no benefit. Extending this analysis to larger systems, other problem classes, and noisy hardware will clarify the practical scope of the HQW advantage.

\section{Numerical Experiments}
\label{num.exp.}
To benchmark the performance of HQW framework, we conduct numerical experiments on two canonical combinatorial optimization problems: Max-Cut and Maximum Independent Set (MIS). Our primary objectives are (i) to compare HQW against standard QAOA across diverse problem instances, (ii) to isolate the contribution of the coin operator and path superposition to any observed advantage, and (iii) to establish a connection between performance and the algebraic measures of Hamiltonian incompatibility discussed in Sec.~\ref{DLA}. All experiments are implemented using PennyLane v0.34.0 \cite{49} with NumPy v1.24.3 \cite{50}. Quantum simulations execute on the \texttt{default.qubit} simulator. Visualization is performed using Matplotlib v3.7.1 \cite{51}, and data analysis is performed using Pandas v2.0.3 \cite{52}. The dataset and analysis scripts are available at \cite{59}. Further implementation details, including circuit diagrams and optimization protocols, are provided in ~\ref{app:num_exp}.

\subsection{Performance on Max-Cut: Benchmark on 50 Random Graphs}

We first compare the average performance of HQW and QAOA over 100 random initializations on a set of 50 8-vertex connected graphs. This metric reflects the typical behavior of each algorithm. For each instance, we perform 100 random initializations of variational parameters, each sampled uniformly from $[0,2\pi]$, and run 300 optimization steps using the Adam optimizer with learning rate $0.1$. For QAOA we fix the depth to $p=2$, and for HQW we set the number of steps to $2p=4$, so that both circuits apply the same total number of Hamiltonian evolution operations. For each graph and each algorithm, we compute the average $1-r$ (gap from optimal approximation ratio $r=1$) over all initializations. 

\begin{figure}[htbp]
\centering
\subfigure[]{
\begin{minipage}[t]{0.45\linewidth}
\centering
\includegraphics[width=0.9\textwidth]{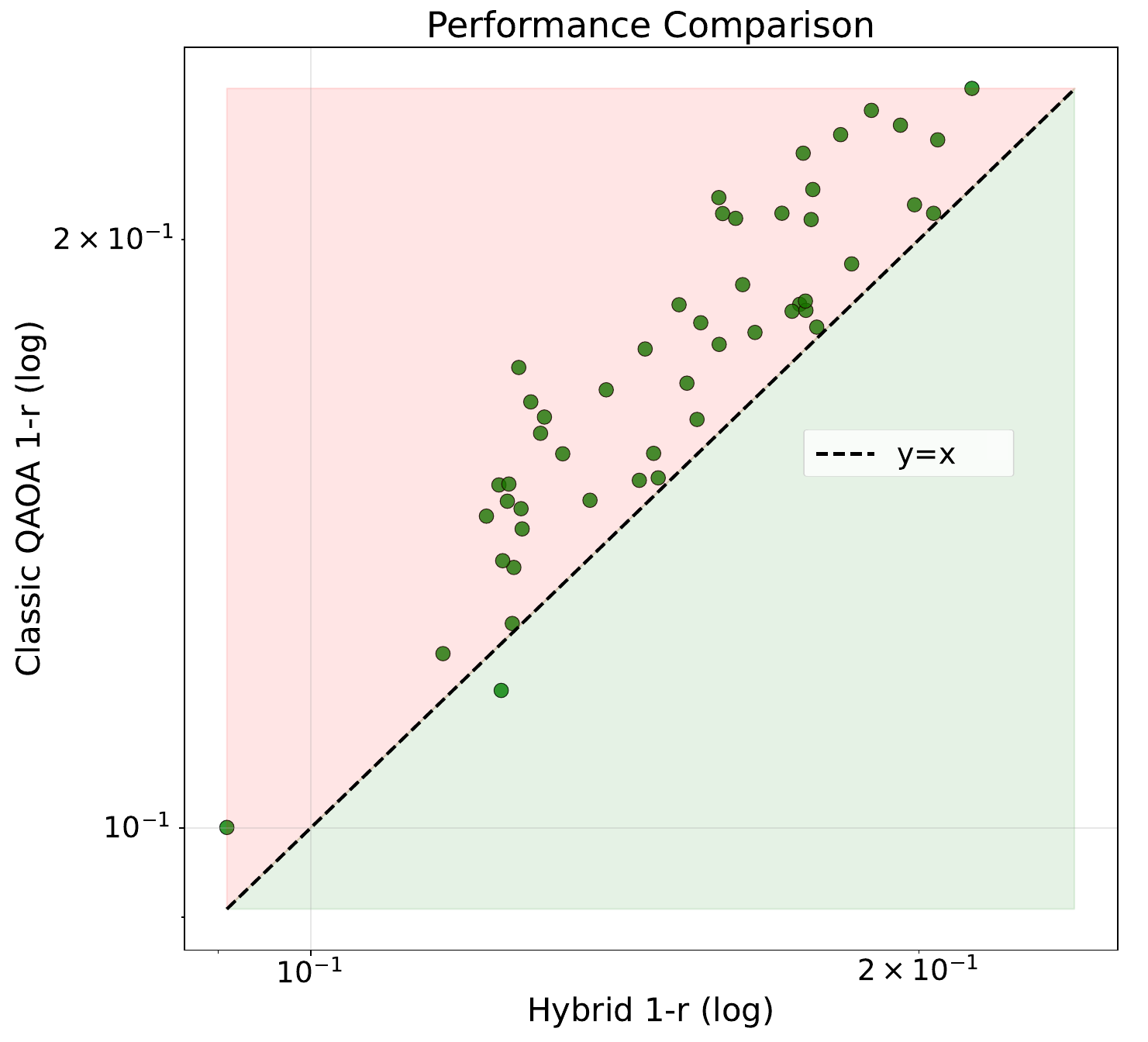}
\label{fig:13}
\end{minipage}
}
\subfigure[]{
\begin{minipage}[t]{0.45\linewidth}
\centering
\includegraphics[width=0.9\textwidth]{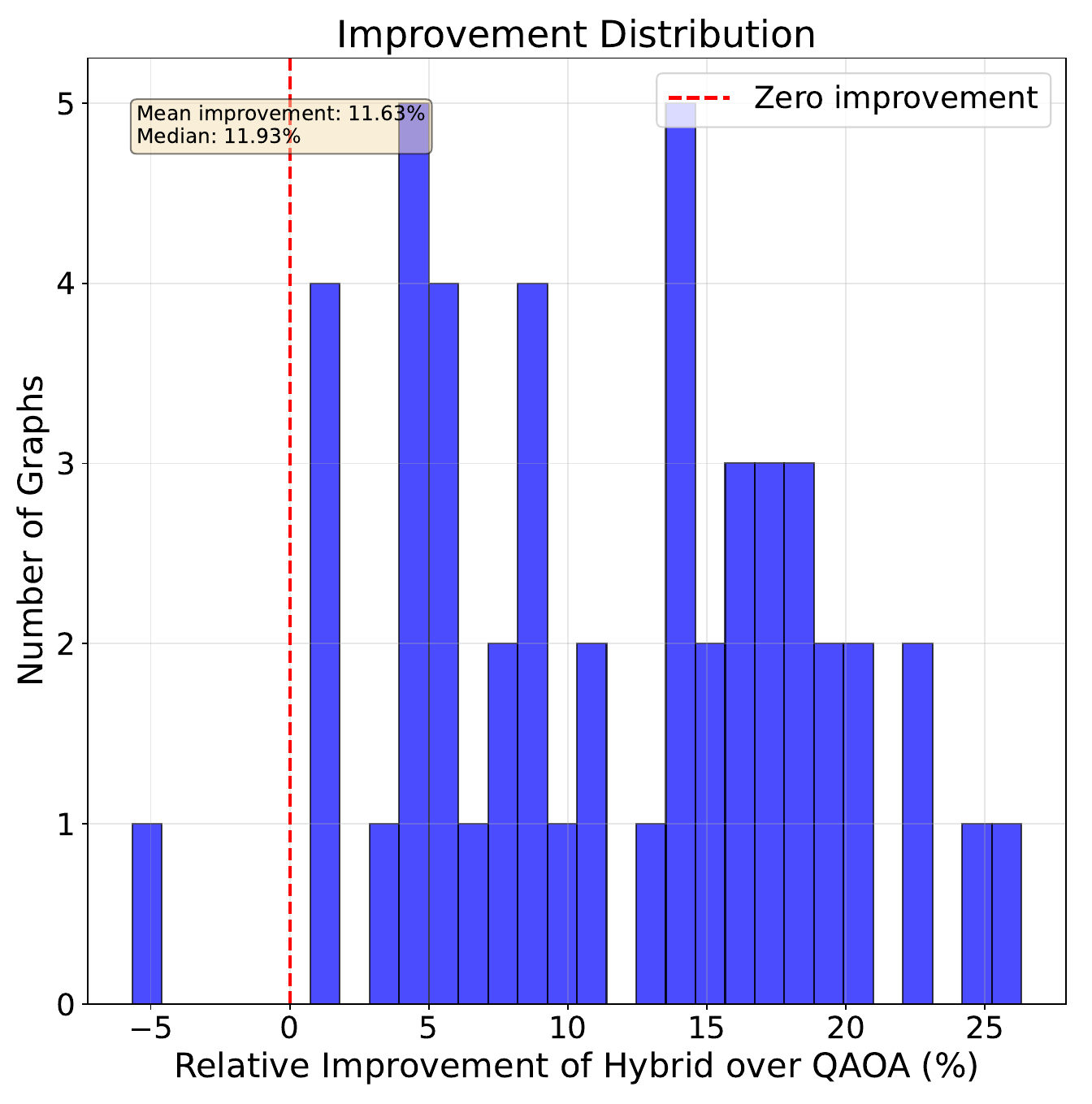}
\label{fig:14}
\end{minipage}
}
\centering
\caption{(a) Scatter plot of average $1-r$ over 100 initializations for QAOA vs. HQW on 50 8-vertex graphs with 18-23 edges. Each point corresponds to one graph. The diagonal line indicates equal performance. Both axes show $1-r$ on a logarithmic scale. (b) Distribution of HQW's relative improvement over QAOA across 50 8-vertex graphs with 18-23 edges for the average-performance comparison. Mean: 11.63\%, median: 11.93\%.} 
\end{figure}

For 50 8-vertex graphs with 18--23 edges, Table~\ref{tab:avg_main} summarizes the results. HQW achieves a lower average $1-r$ ($0.154200$) than QAOA ($0.175399$). HQW yields better average performance in 49 out of 50 instances, giving HQW a win rate of $98\%$. The average relative improvement is $11.63\%$, and the median relative improvement is $11.93\%$. A scatter plot comparing the average $1-r$ values for each graph is shown in FIG.~\ref{fig:13}. All points lie on or above the $y=x$ line except for a single point where QAOA slightly outperforms HQW. The distribution of relative improvements (FIG.~\ref{fig:14}) shows that the majority of instances favor HQW, with a clear concentration of improvements between $5\%$ and $20\%$. HQW thus gives more reliable average-case performance than standard QAOA on these graphs.

\begin{table}[htbp]
\centering
\begin{tabular}{lcc}
\toprule
\textbf{Metric} & \textbf{QAOA} & \textbf{HQW} \\
\hline
Average $1-r$ value & 0.175399 & 0.154200 \\
Graphs with better performance & 1 & 49 \\
Win rate & 2\% & 98\% \\
Average relative improvement & -- & +11.63\% \\
Median relative improvement & -- & +11.93\% \\
\midrule
\end{tabular}
\caption{Comparison of average performance over 100 random initializations on a set of 50 8-vertex graphs with 18-23 edges. Lower $1-r$ is better.}
\label{tab:avg_main}
\end{table}

\begin{figure}[htbp]
\centering
\subfigure[]{
\begin{minipage}[t]{0.45\linewidth}
\centering
\includegraphics[width=0.9\textwidth]{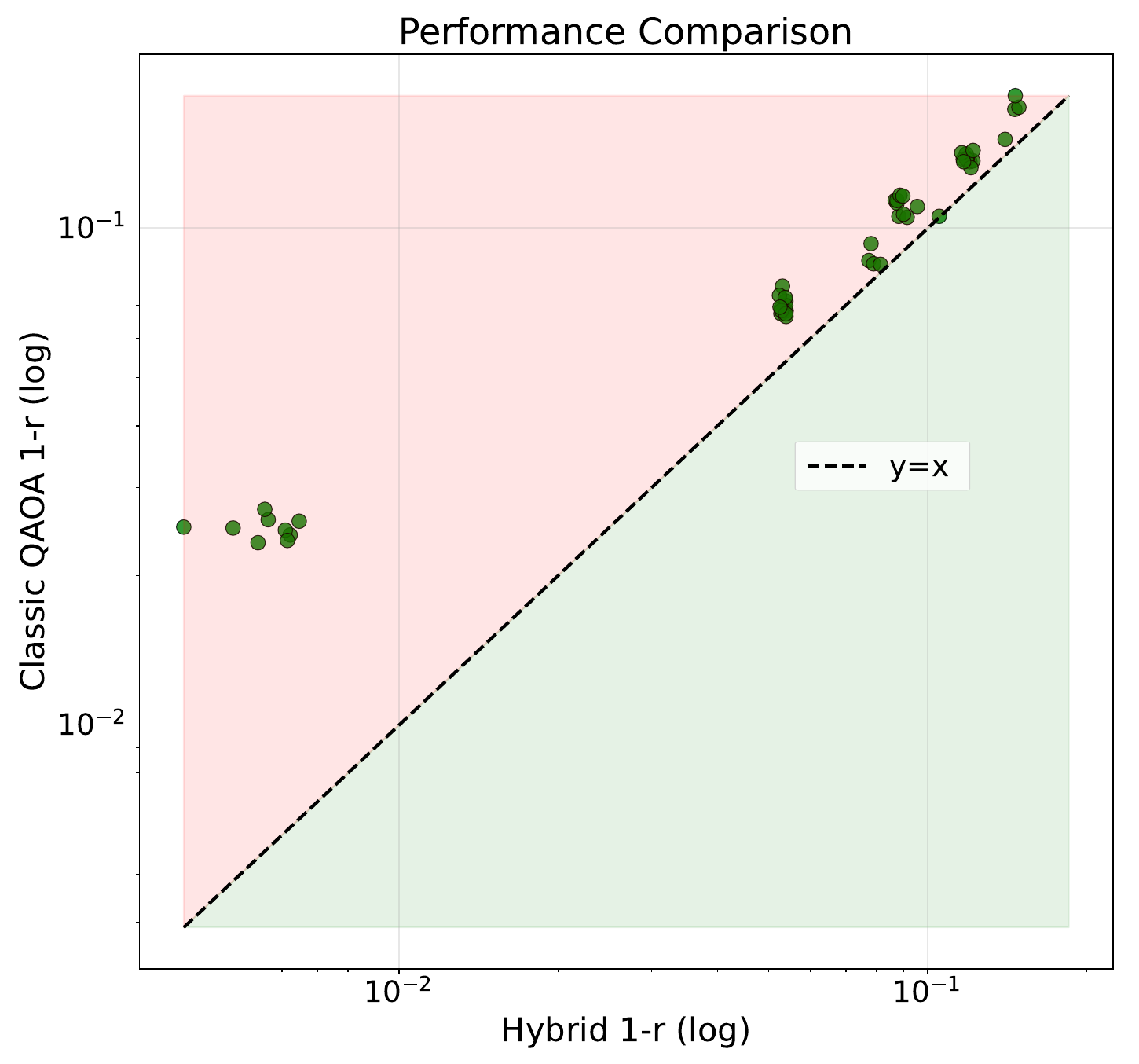}
\label{fig:15}
\end{minipage}
}
\subfigure[]{
\begin{minipage}[t]{0.45\linewidth}
\centering
\includegraphics[width=0.9\textwidth]{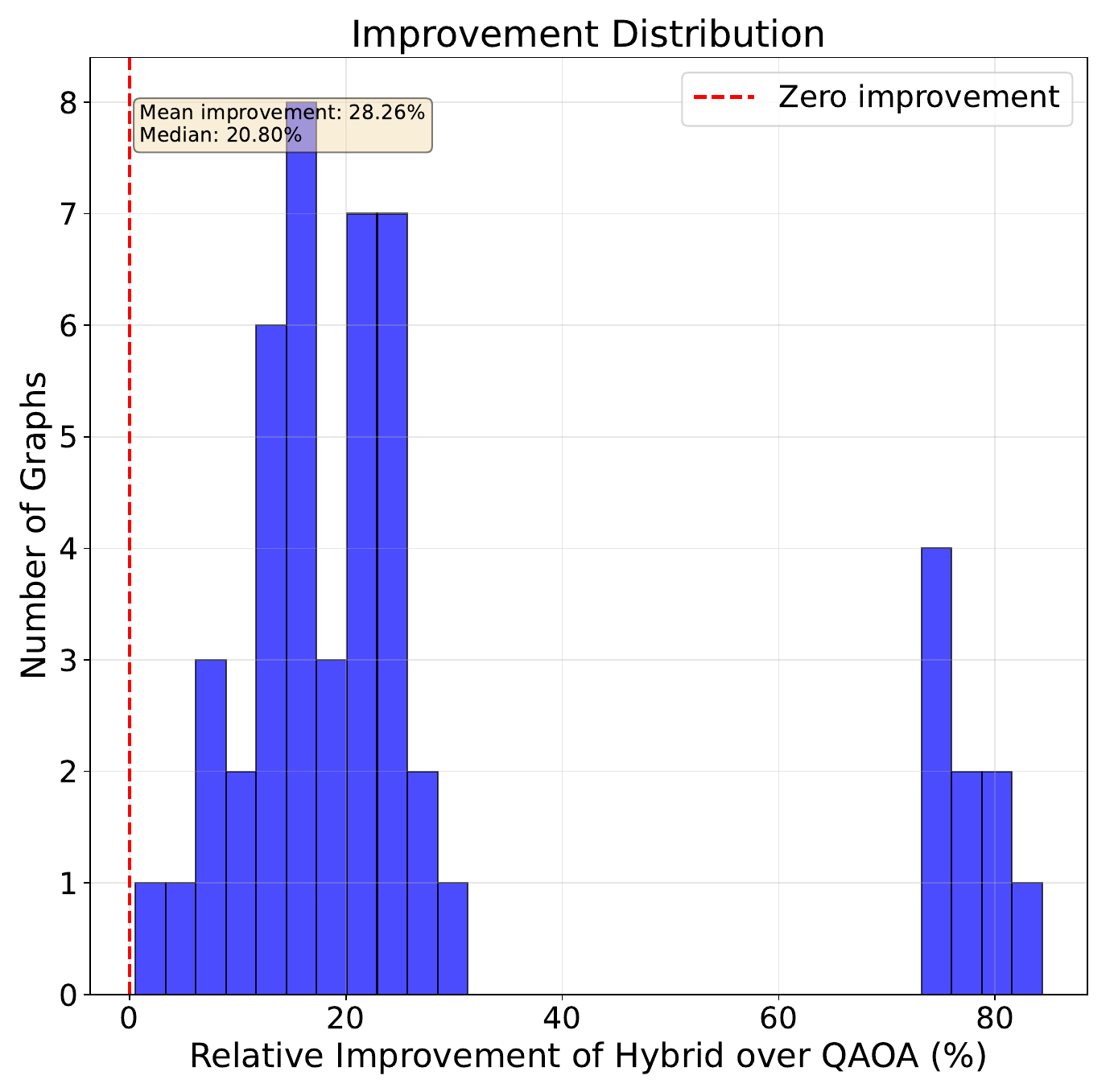}
\label{fig:16}
\end{minipage}
}
\centering
\caption{(a) Scatter plot of average $1-r$ over 100 initializations for QAOA vs. HQW on 50 8-vertex graphs with 24-28 edges. Each point corresponds to one graph. The diagonal line indicates equal performance. Both axes show $1-r$ on a logarithmic scale. (b) Distribution of HQW's relative improvement over QAOA across 50 8-vertex graphs with 24-28 edges for the average-performance comparison. Mean: 28.26\%, median: 20.80\%.} 
\end{figure}

For 50 8-vertex graphs with 24--28 edges, Table~\ref{tab:avg_main2} summarizes the results. HQW achieves a lower average $1-r$ ($0.074771$) than QAOA ($0.093015$). HQW yields better average performance in all of 50 instances. The average relative improvement is $28.26\%$, and the median relative improvement is $20.80\%$. The scatter and distribution plots are shown in FIG.~\ref{fig:15} and FIG.~\ref{fig:16}. From the results presented in Table~\ref{tab:avg_main} and Table~\ref{tab:avg_main2}, we can observe that as the edge density of the graphs increases, the performance improvement of HQW relative to QAOA becomes more significant, so HQW is more effective on denser graphs. 
\begin{table}[htbp]
\centering
\begin{tabular}{lcc}
\toprule
\textbf{Metric} & \textbf{QAOA} & \textbf{HQW} \\
\hline
Average $1-r$ value & 0.093015 & 0.074771 \\
Graphs with better performance & 0 & 50 \\
Win rate & 0\% & 100\% \\
Average relative improvement & -- & +28.26\% \\
Median relative improvement & -- & +20.80\% \\
\hline
\end{tabular}
\caption{Comparison of average performance over 100 random initializations on a set of 50 8-vertex graphs with 24-28 edges. Lower $1-r$ is better.}
\label{tab:avg_main2}
\end{table}

\begin{figure}[htbp]
\centering
\subfigure[]{
\begin{minipage}[t]{0.45\linewidth}
\centering
\includegraphics[width=0.9\textwidth]{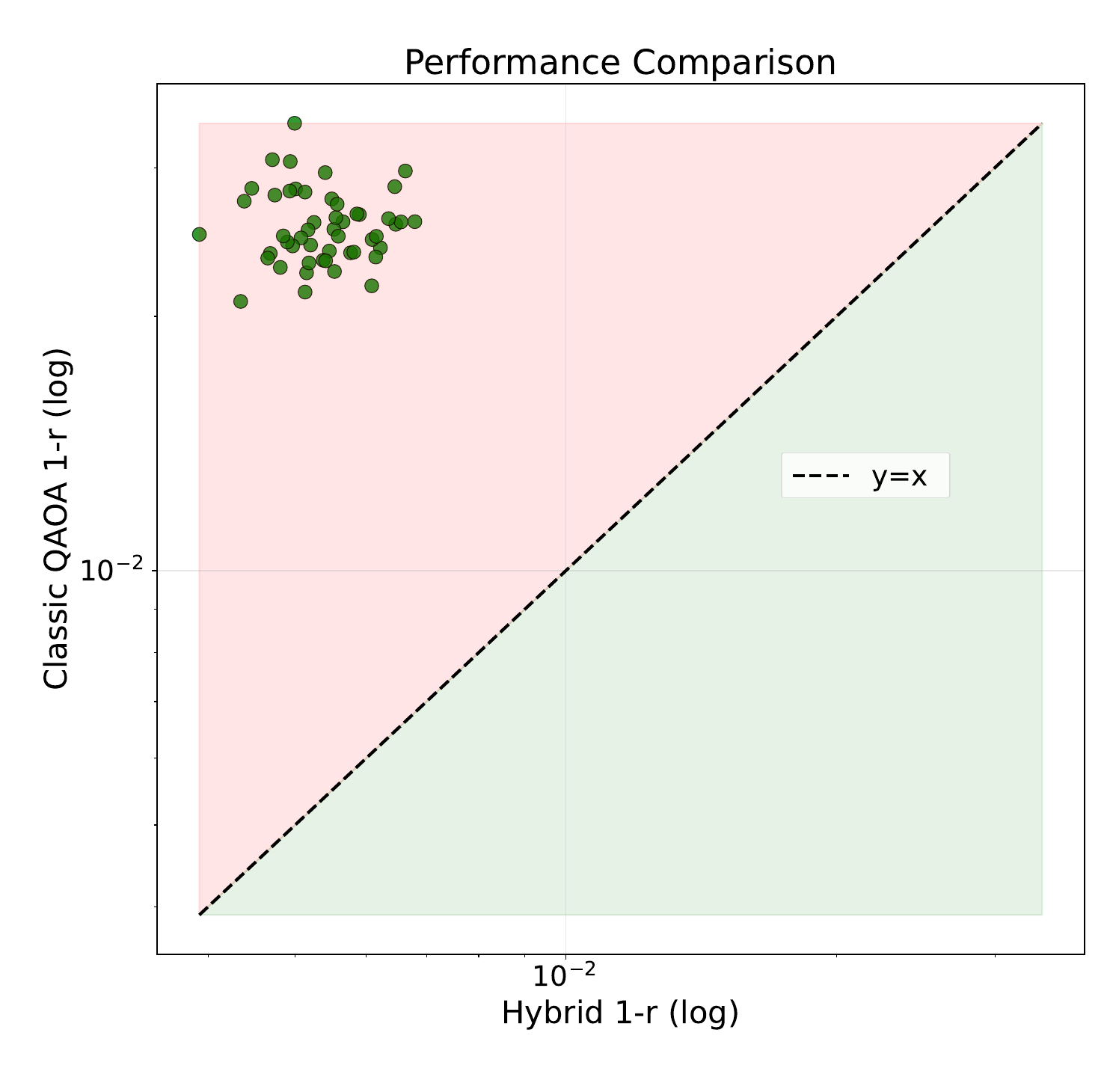}
\label{fig:17}
\end{minipage}
}
\subfigure[]{
\begin{minipage}[t]{0.45\linewidth}
\centering
\includegraphics[width=0.9\textwidth]{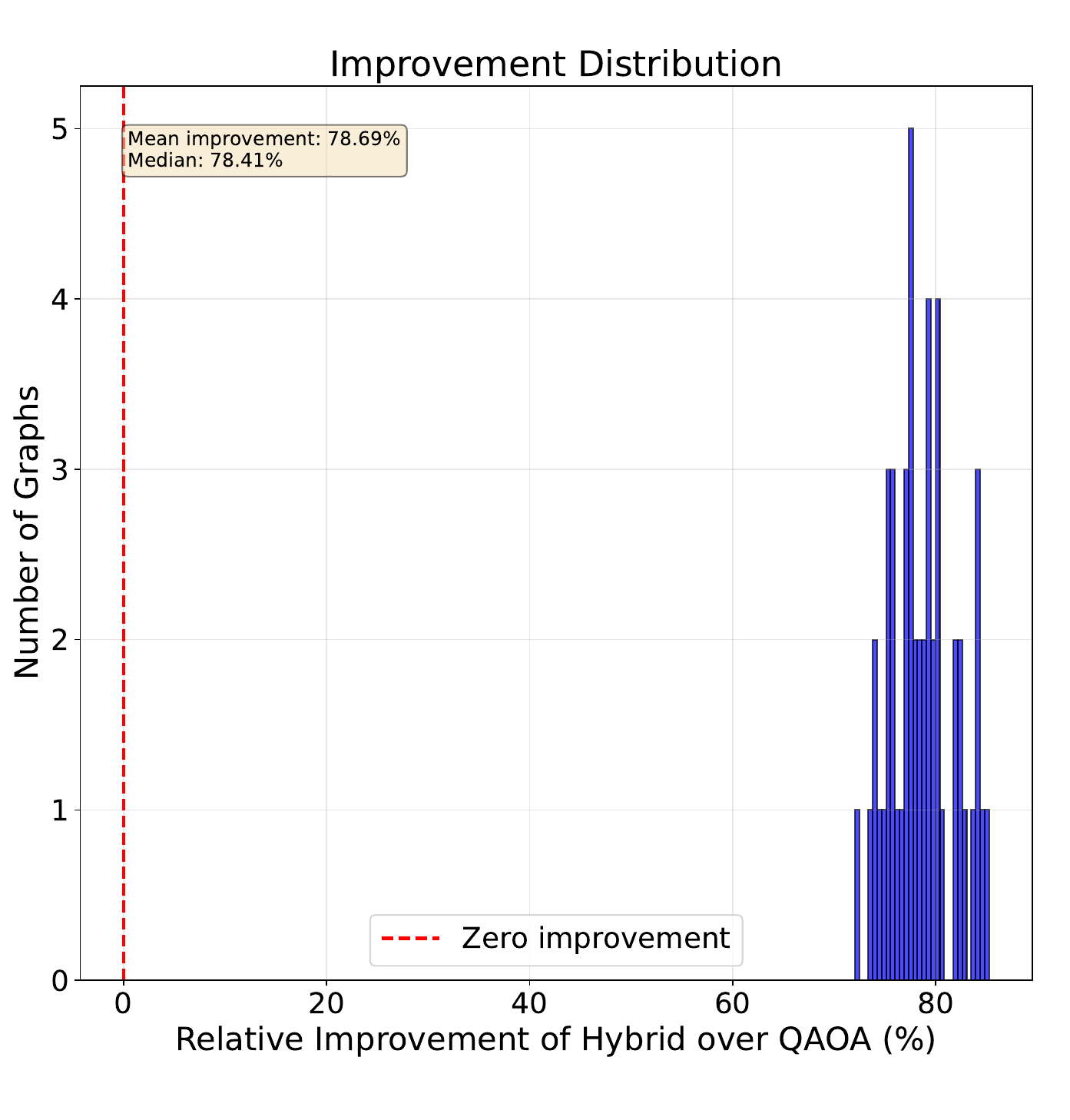}
\label{fig:18}
\end{minipage}
}
\centering
\caption{(a) Scatter plot of average $1-r$ over 100 initializations for QAOA vs. HQW on 50 8-vertex complete graphs. Each point corresponds to one graph. The diagonal line indicates equal performance. Both axes show $1-r$ on a logarithmic scale. (b) Distribution of HQW's relative improvement over QAOA across 50 8-vertex complete graphs for the average-performance comparison. Mean: 78.69\%, median: 78.41\%.} 
\end{figure}

Meanwhile, from FIG.~\ref{fig:15} and FIG.~\ref{fig:16}, we find that the performance gain of HQW over QAOA can reach approximately $80\%$ on certain graphs. During our experiments, these graphs were identified as the complete graph with 8 vertices. Therefore, we also conducted comparative experiments on the 8-vertex complete graph $K_8$ (50 random initializations). HQW achieves better average performance on $K_8$. The average performance gain reaches 78.69\%, with a median improvement of 78.41\%. The scatter and distribution plots are shown in FIG.~\ref{fig:17} and FIG.~\ref{fig:18}. HQW therefore holds a large performance advantage for Max-Cut on complete graphs.

We also compare the best $1-r$ obtained across the 100 initializations for each algorithm, which reflects the optimal achievable performance with sufficient initialization search. On a separate set of 50 8-vertex graphs with 20-28 edges under the same protocol, HQW matches or exceeds QAOA in all instances under a tolerance of $10^{-6}$. Table~\ref{tab:1} summarizes the results. HQW achieves a better average approximation ratio than QAOA in 68\% of the instances (34 out of 50) and matches it in the remaining 32\% (16 instances), with an average relative improvement of 0.86\%. These results, detailed in ~\ref{app:num_exp}, confirm that HQW consistently matches or exceeds the performance of standard QAOA for this problem set, despite the modest absolute improvement in approximation ratio.

\begin{table}[htbp]
\centering
\begin{tabular}{lcc}
\toprule
\textbf{Metric} & \textbf{QAOA} & \textbf{HQW} \\
\hline
Average 1-r value & 0.112273 & 0.111423 \\
Graphs with better performance & 0 & 34 \\
Win rate & 0\% & 68\% \\
Average relative improvement & -- & +0.86\% \\
Median relative improvement & -- & +0.36\% \\
\hline
\end{tabular}
\caption{Comparison of best-performance over 100 random initializations on a set of 50 random 8-vertex Max-Cut instances.}
\label{tab:1}
\end{table}

\subsection{Component Analysis: Role of the Coin Operator}

\begin{figure}[htbp]
\centering
\subfigure[]{
\begin{minipage}[t]{0.45\linewidth}
\centering
\includegraphics[width=0.9\textwidth]{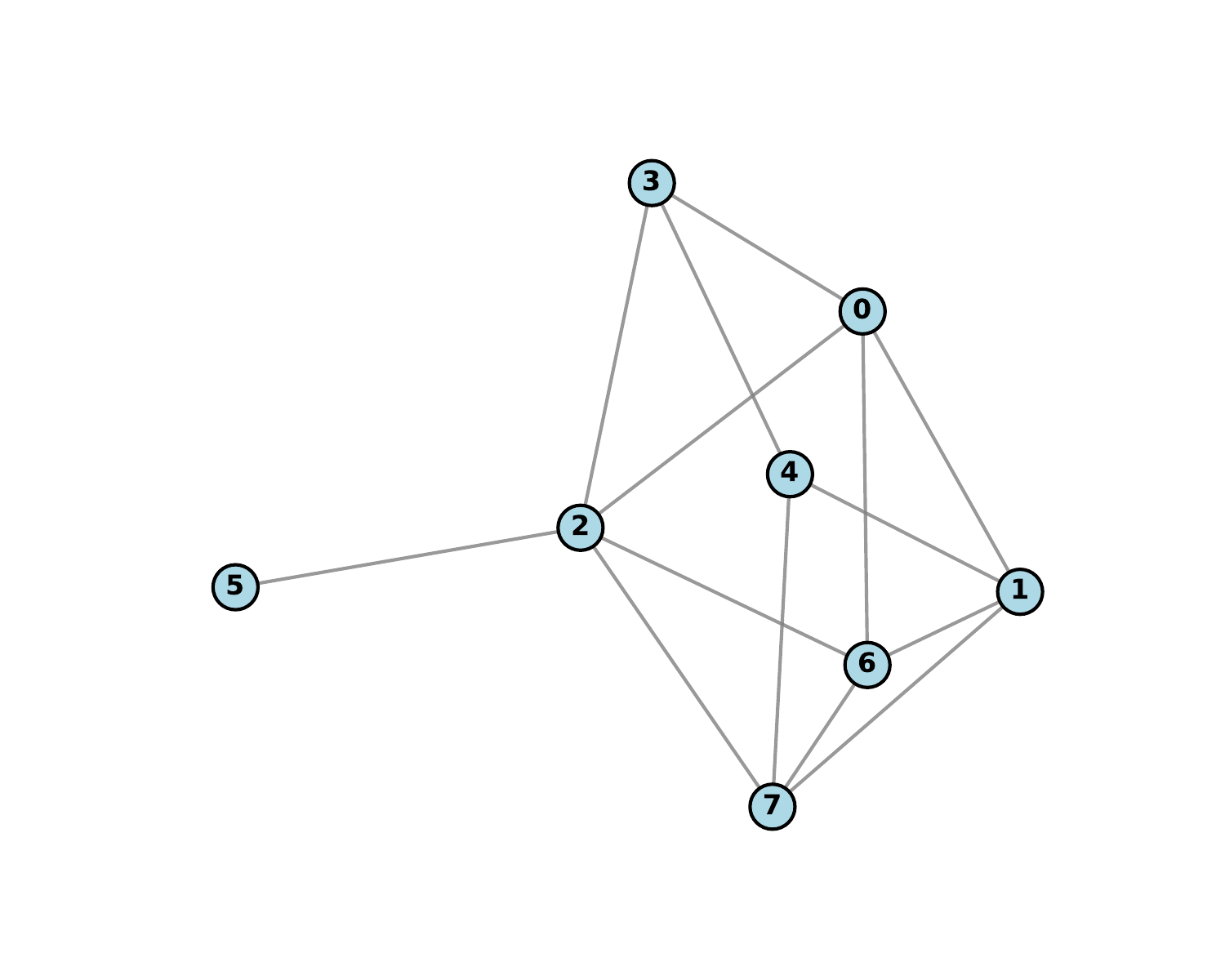}
\label{fig:s5}
\end{minipage}
}
\subfigure[]{
\begin{minipage}[t]{0.45\linewidth}
\centering
\includegraphics[width=0.9\textwidth]{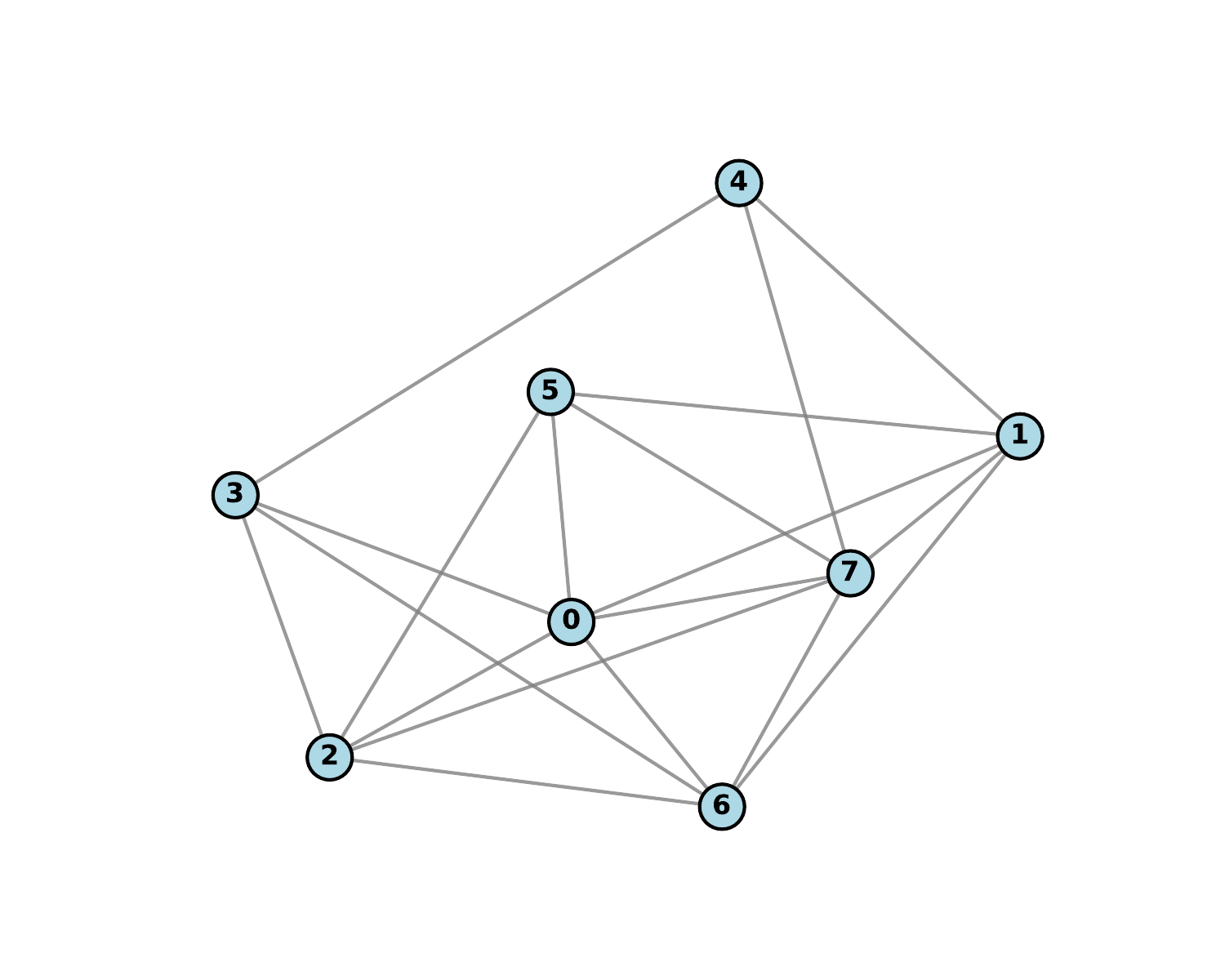}
\label{fig:s6}
\end{minipage}
}
\centering
\caption{(a) The 8-vertex graph for the Max-Cut problem. (b) The 8-vertex graph for the Maximum Independent Set problem.} 
\end{figure}

To isolate the contribution of the coin operator and path superposition, we compare four algorithms on two representative graphs: an 8-vertex, 14-edge graph for Max-Cut (FIG.~\ref{fig:s5}) and an 8-vertex, 19-edge graph for Maximum Independent Set (FIG.~\ref{fig:s6}). 

The four algorithms are:
\begin{enumerate}
\item \textbf{Standard QAOA}: The standard QAOA ansatz with $p$ layers, each consisting of $e^{-i\gamma H_c}e^{-i\beta H_b}$.
\item \textbf{HQW}: The hybrid quantum walk ansatz with $2p$ steps, incorporating a coin qubit and controlled evolutions.
\item \textbf{QAOA Variant 1}: A QAOA circuit where the parameters $\gamma_k$ (for $H_c$) are taken from the odd steps of the optimized HQW circuit, and $\beta_k$ (for $H_b$) from the even steps.
\item \textbf{QAOA Variant 2}: The opposite interleaving: $\gamma_k$ from even steps and $\beta_k$ from odd steps.
\end{enumerate}
These two variants allow us to assess whether any advantage of HQW can be replicated simply by reordering the QAOA parameters, or whether it requires the explicit coin-controlled structure.

\begin{figure}[htbp]
\centering
\subfigure[]{
\begin{minipage}[t]{0.9\linewidth}
\centering
\includegraphics[width=\textwidth]{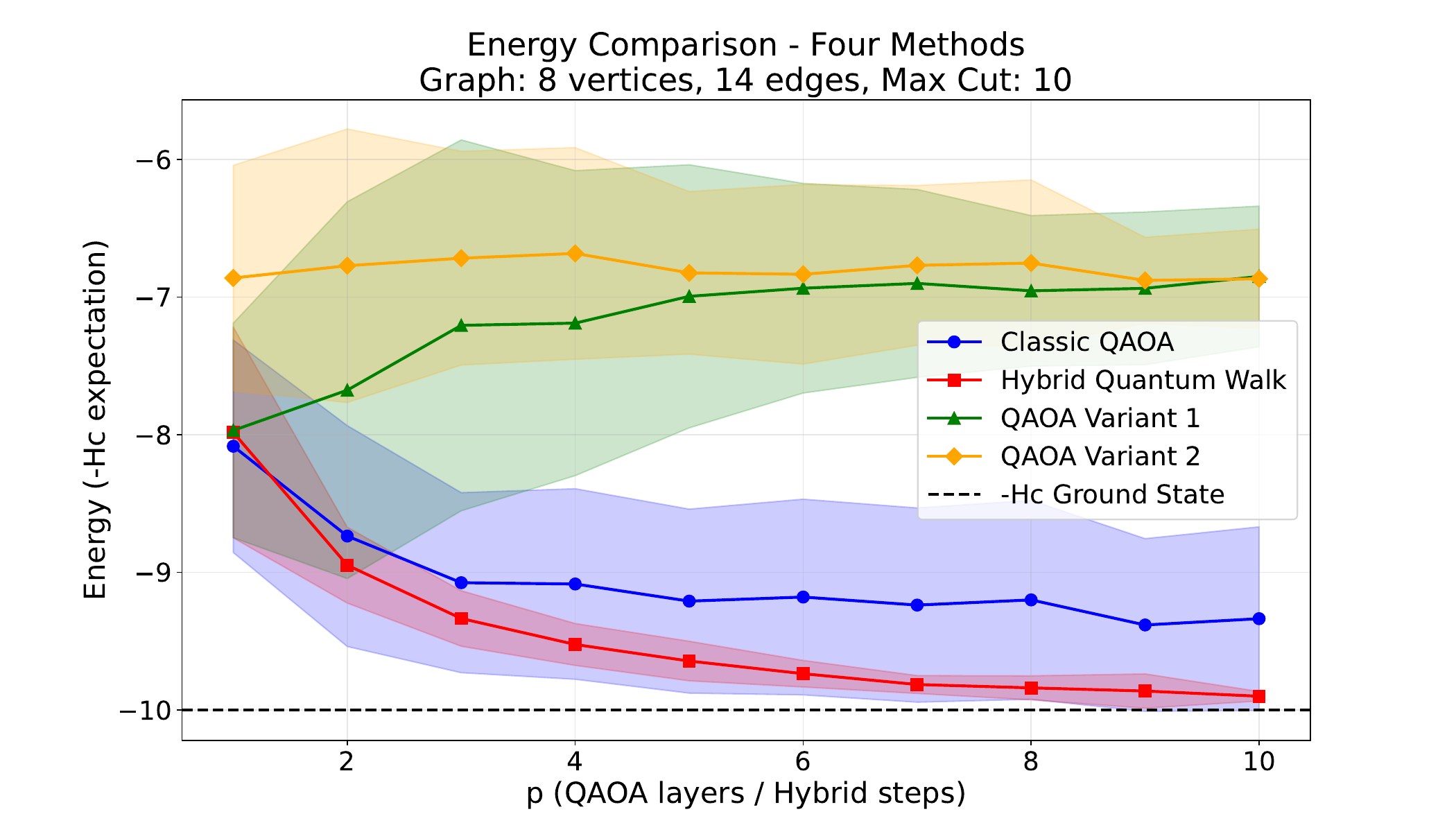}
\label{fig:1}
\end{minipage}
}

\subfigure[]{
\begin{minipage}[t]{0.9\linewidth}
\centering
\includegraphics[width=\textwidth]{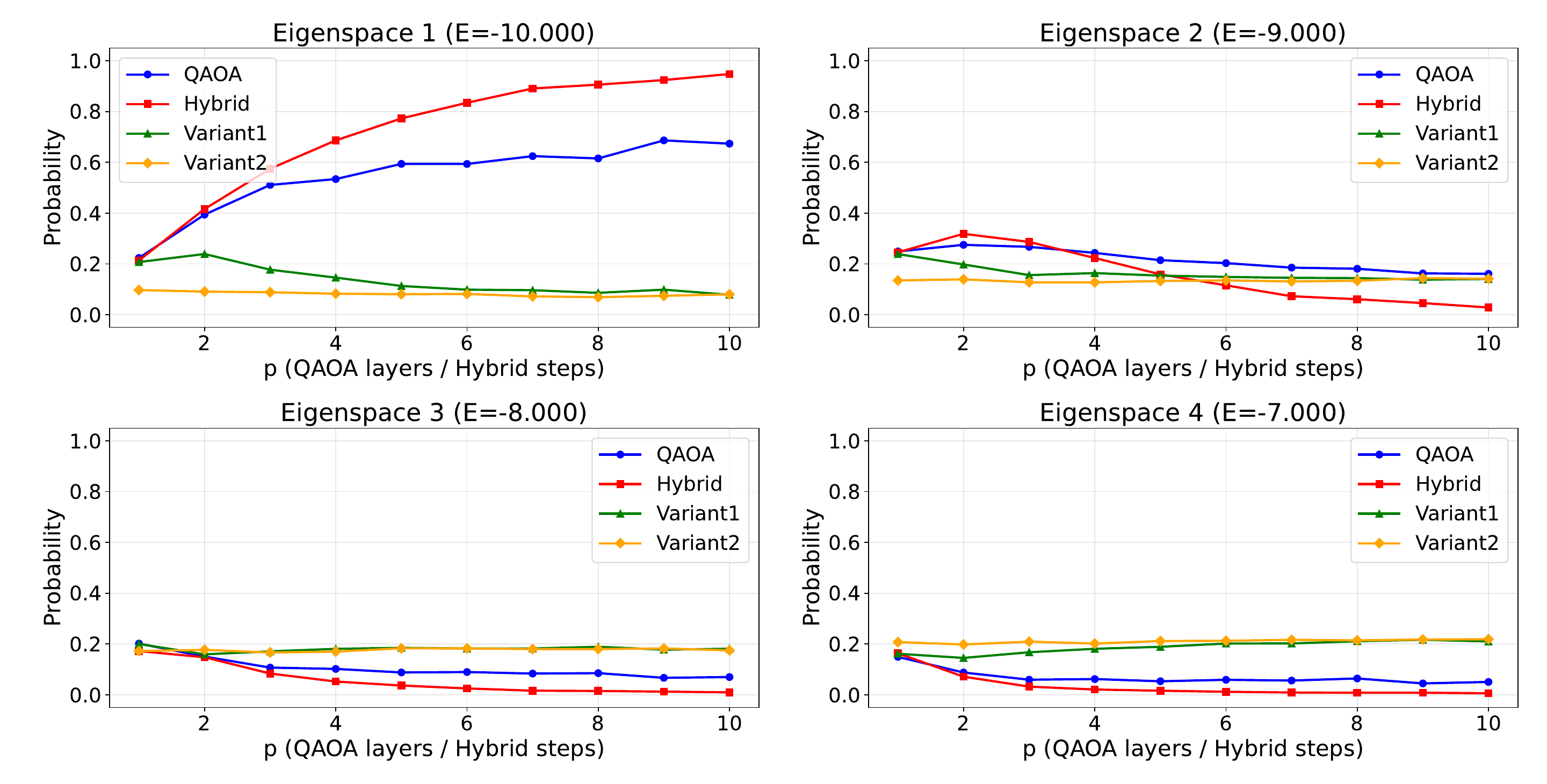}
\label{fig:2}
\end{minipage}
}
\centering
\caption{Experimental results on an 8-vertex Max-Cut instance. (a) Expectation value of $-H_C$ as a function of depth $p$ for QAOA or $2p$ steps for HQW. Shaded regions indicate standard deviation over 100 random initializations. The dashed black line marks the ground state energy. (b) Projection probabilities onto the first four eigenspaces of $-H_C$, aggregated over all initializations. HQW shows the fastest growth in ground state probability with depth.} 
\label{fig:12}
\end{figure}

For each algorithm and depth $p$ ranging from 1 to 10, with $2p$ steps for HQW, we perform 100 random initializations of variational parameters, each optimized for 300 steps using Adam. The results for the Max-Cut problem are shown in FIG.~\ref{fig:12}. FIG.~\ref{fig:12}a plots the expectation value of $-H_c$ as a function of $p$. HQW converges faster and achieves lower values than both classic QAOA and the parameter-extracted variants across all depths. HQW also has the smallest standard deviation among the four methods, showing better robustness to initialization. FIG.~\ref{fig:12}b analyzes the final quantum states by projecting onto the four lowest eigenspaces of $-H_c$ with energies $E_1=-10$, $E_2=-9$, $E_3=-8$, $E_4=-7$. HQW yields the highest probability of occupying the ground state eigenspace, particularly at larger $p$, confirming that its improved performance translates into higher-quality solutions.

\begin{figure}[htbp]
\centering
\subfigure[]{
\begin{minipage}[t]{0.9\linewidth}
\centering
\includegraphics[width=\textwidth]{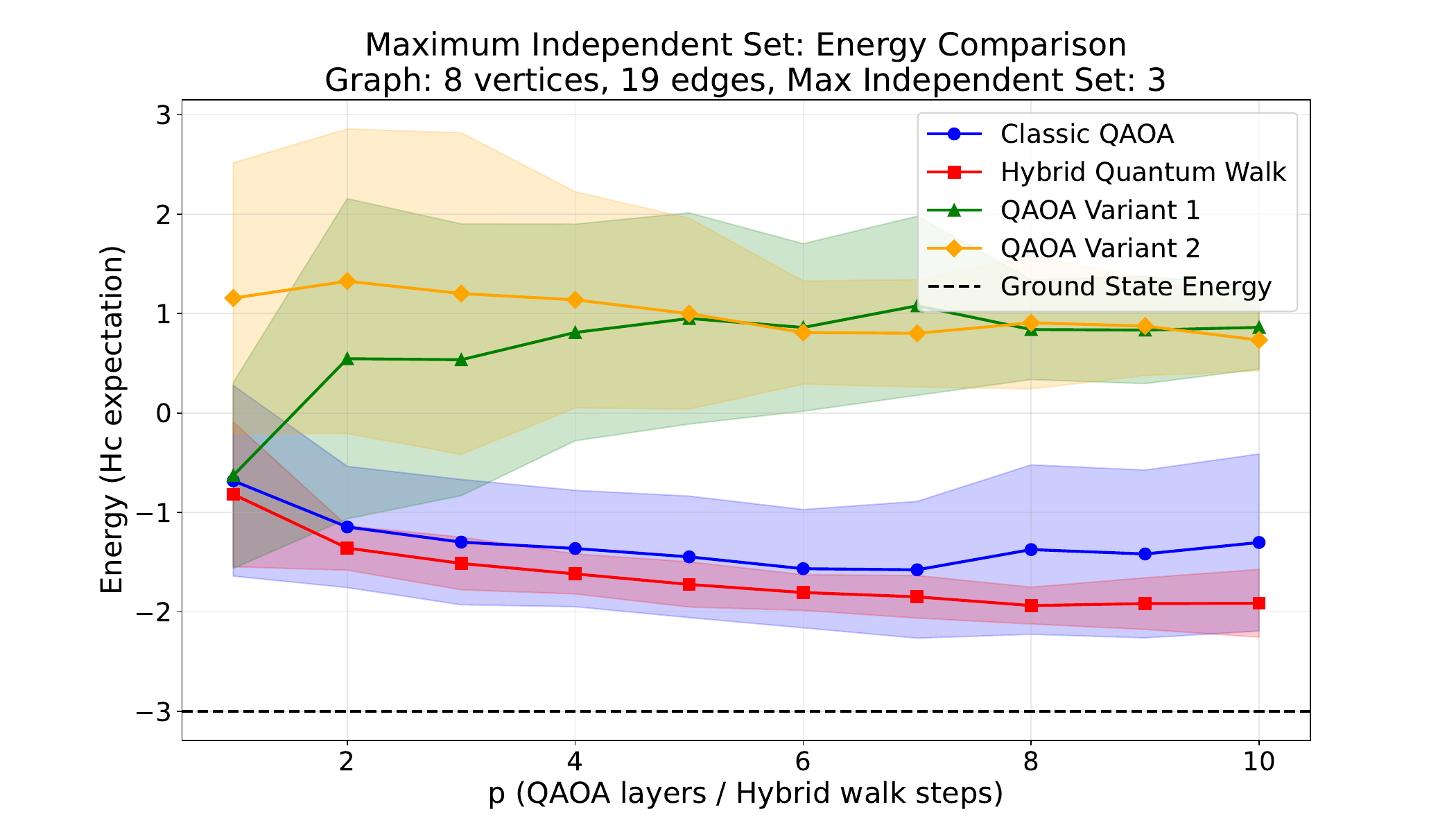}
\label{fig:9}
\end{minipage}
}
\subfigure[]{
\begin{minipage}[t]{0.9\linewidth}
\centering
\includegraphics[width=\textwidth]{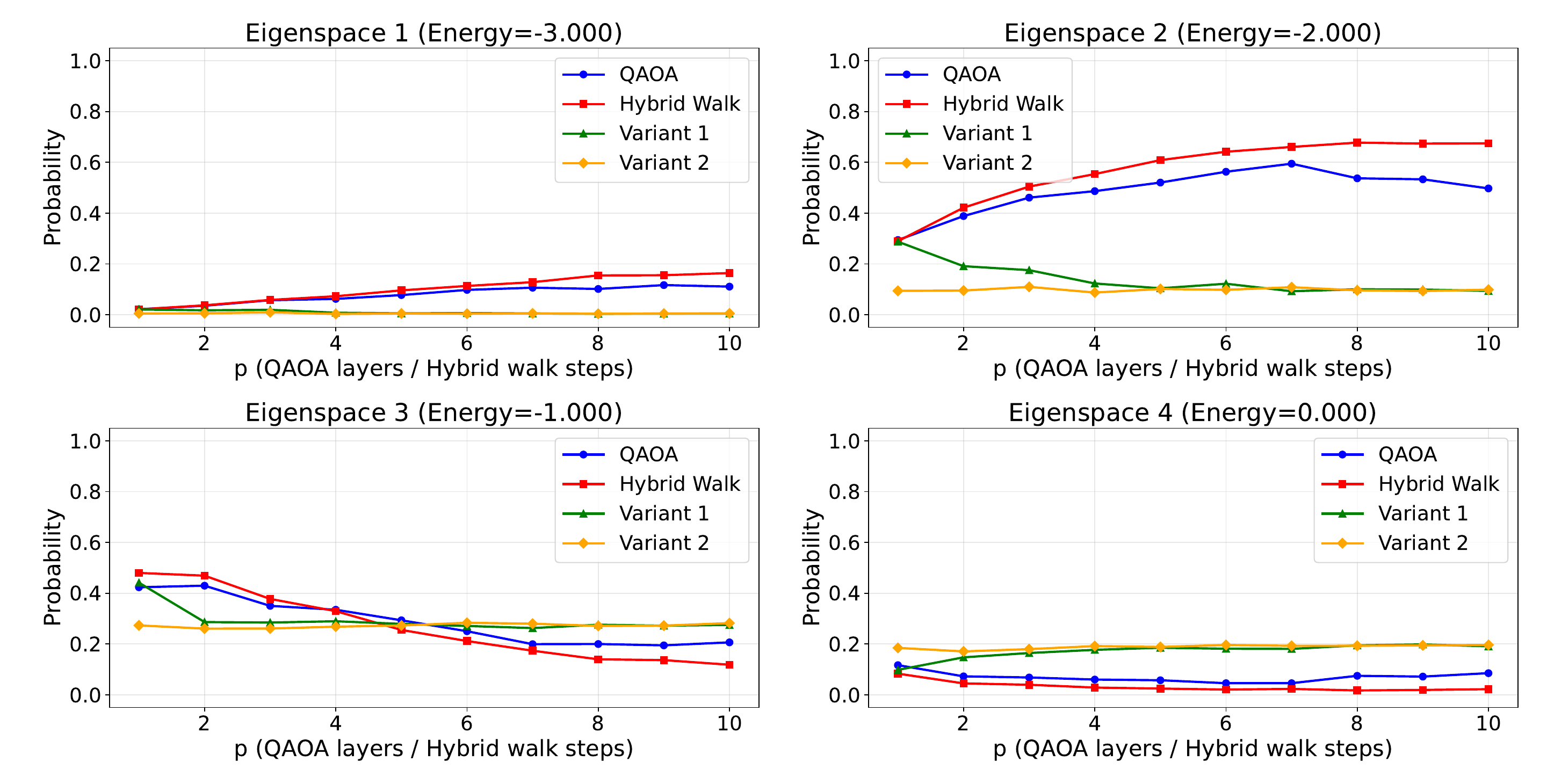}
\label{fig:10}
\end{minipage}
}
\centering
\caption{Experimental results on an 8-vertex MIS instance. (a) Expectation value of $H_C$ as a function of depth. Shaded regions indicate standard deviation over 100 random initializations. The dashed black line marks the ground state energy. (b) Projection probabilities onto the first four eigenspaces of $H_C$ ($E_1=-3$, $E_2=-2$, $E_3=-1$, $E_4=0$).} 
\label{fig:910}
\end{figure}

Analogous results for the Maximum Independent Set problem on the 8-vertex graph are shown in FIG.~\ref{fig:910}. The performance order remains HQW $>$ QAOA $>$ Variant 1/Variant 2, and HQW again achieves the lowest energy and highest ground state probability with the smallest variance.

\subsection{Connection to Jordan Product Negativity}

The performance gains reported above are consistent with the algebraic analysis of Sec.~\ref{DLA}. The Jordan product negativity $|\mathcal{N}_{\min}| > 0$ holds for all benchmarked instances, and Theorem~1 guarantees that HQW's expressive capacity strictly exceeds QAOA's under this condition. The $d_{\rm expl}$ measurements (Fig.~\ref{fig:4}) confirm that this algebraic advantage holds in $97.8\%$ of cases and translates into consistent optimization gains that are independent of graph density. The Jordan-Lie algebra therefore explains why the coin-controlled path-superposition mechanism produces the performance hierarchy HQW $>$ QAOA $>$ parameter-extracted variants.

\subsection{Scalability to Larger Graphs}

\begin{figure}[htbp]
\centering
\subfigure[]{
\begin{minipage}[t]{0.45\linewidth}
\centering
\includegraphics[width=0.9\textwidth]{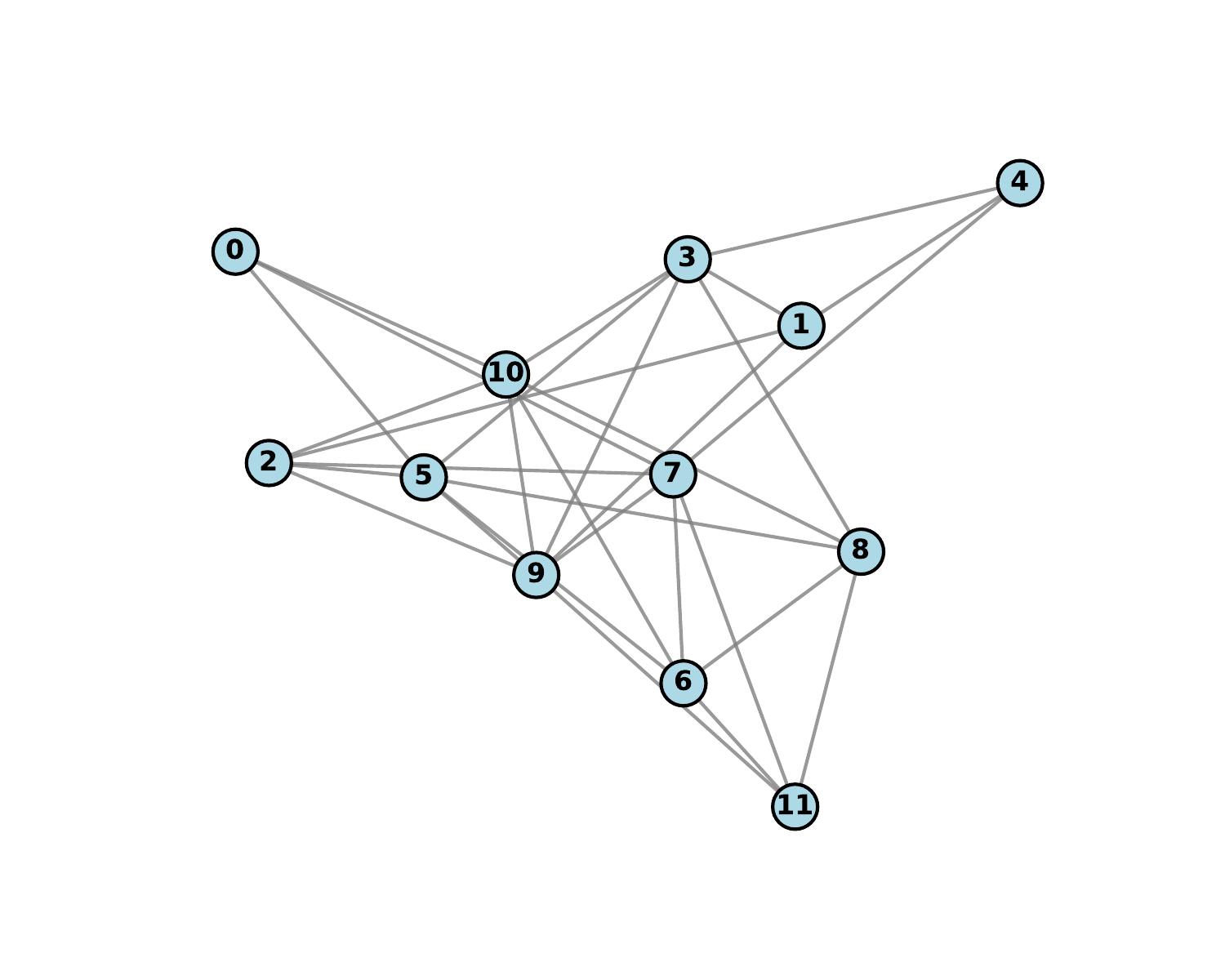}
\label{fig:s7}
\end{minipage}
}
\subfigure[]{
\begin{minipage}[t]{0.45\linewidth}
\centering
\includegraphics[width=0.9\textwidth]{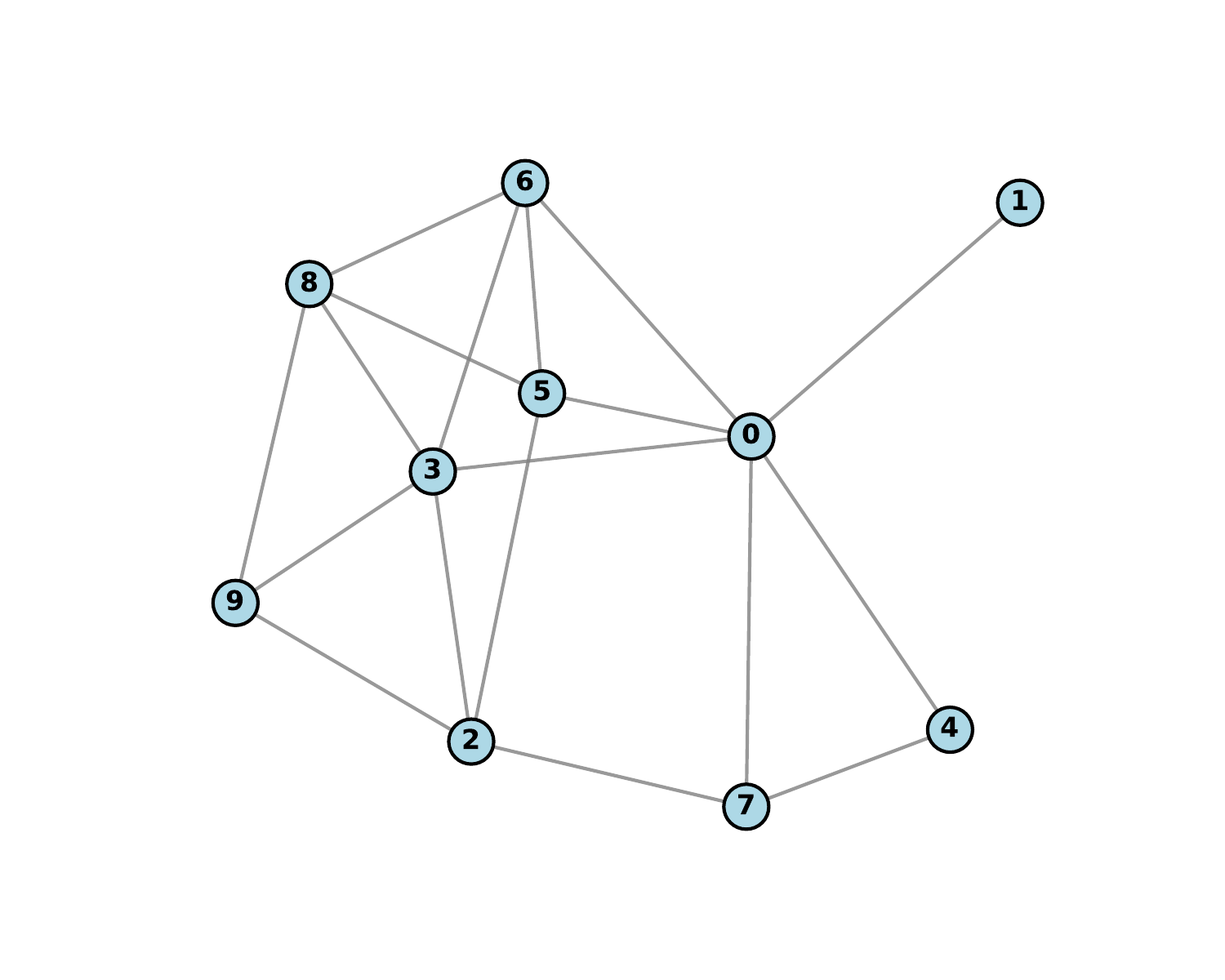}
\label{fig:s8}
\end{minipage}
}
\centering
\caption{(a) The 12-vertex graph for the Max-Cut problem. (b) The 10-vertex graph for the Maximum Independent Set problem.} 
\end{figure}

To assess scalability, we tested HQW and QAOA on larger graphs: a 12-vertex, 30-edge Max-Cut instance (FIG.~\ref{fig:s7}) and a 10-vertex MIS instance (FIG.~\ref{fig:s8}). For these tests, we increased the number of random initializations to 200 and ran 150 optimization steps per initialization to manage computational cost while maintaining statistical significance.

\begin{figure}[htbp]
\centering
\includegraphics[width=0.8\textwidth]{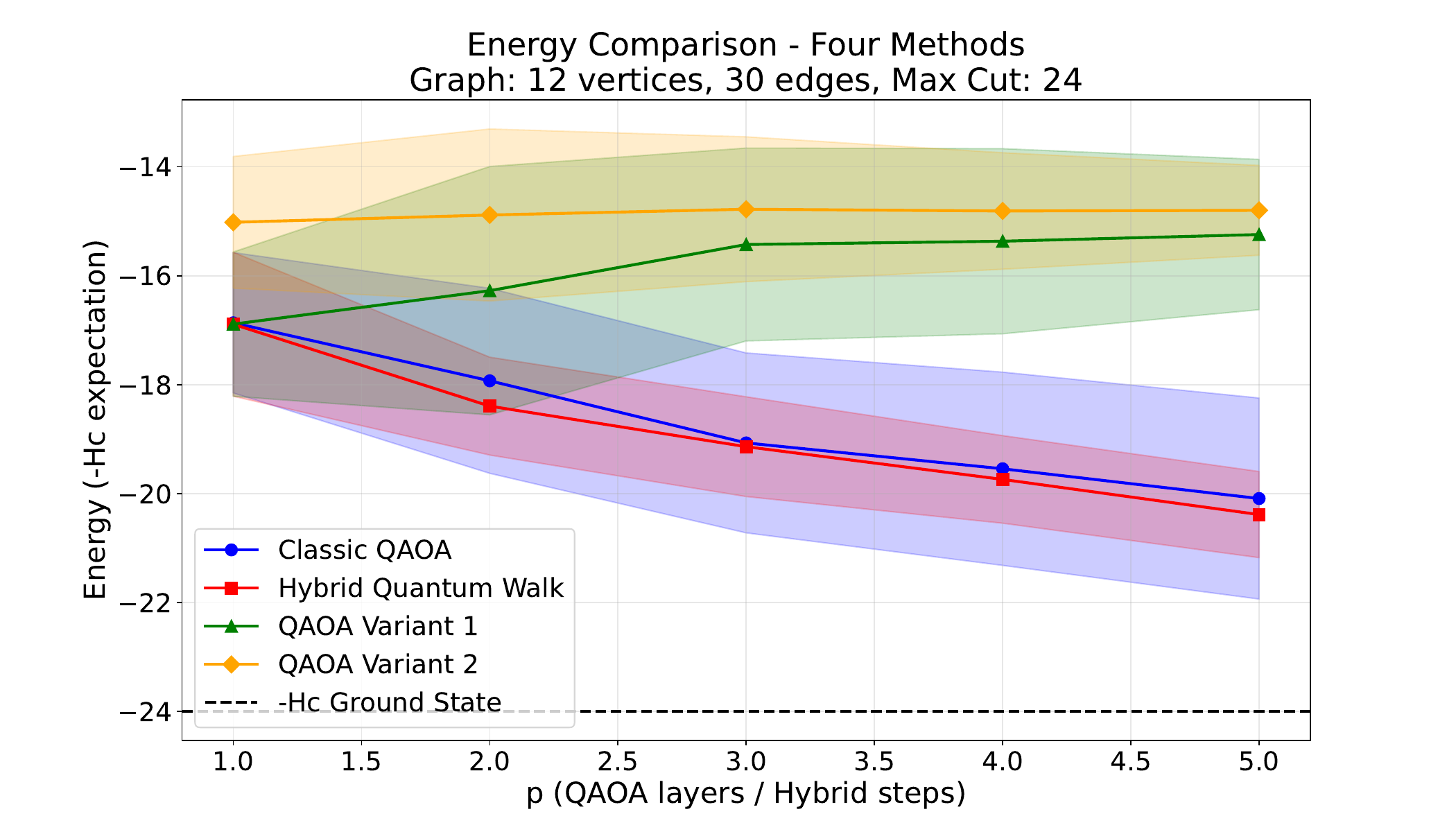}
\caption{Performance comparison on a 12-vertex Max-Cut graph. The vertical axis shows the expectation value of $-H_c$, with ground state energy $-24$. HQW converges faster and achieves lower energy at low depths.}
\label{fig:s9}
\end{figure}

\begin{figure}[htbp]
\centering
\includegraphics[width=0.8\textwidth]{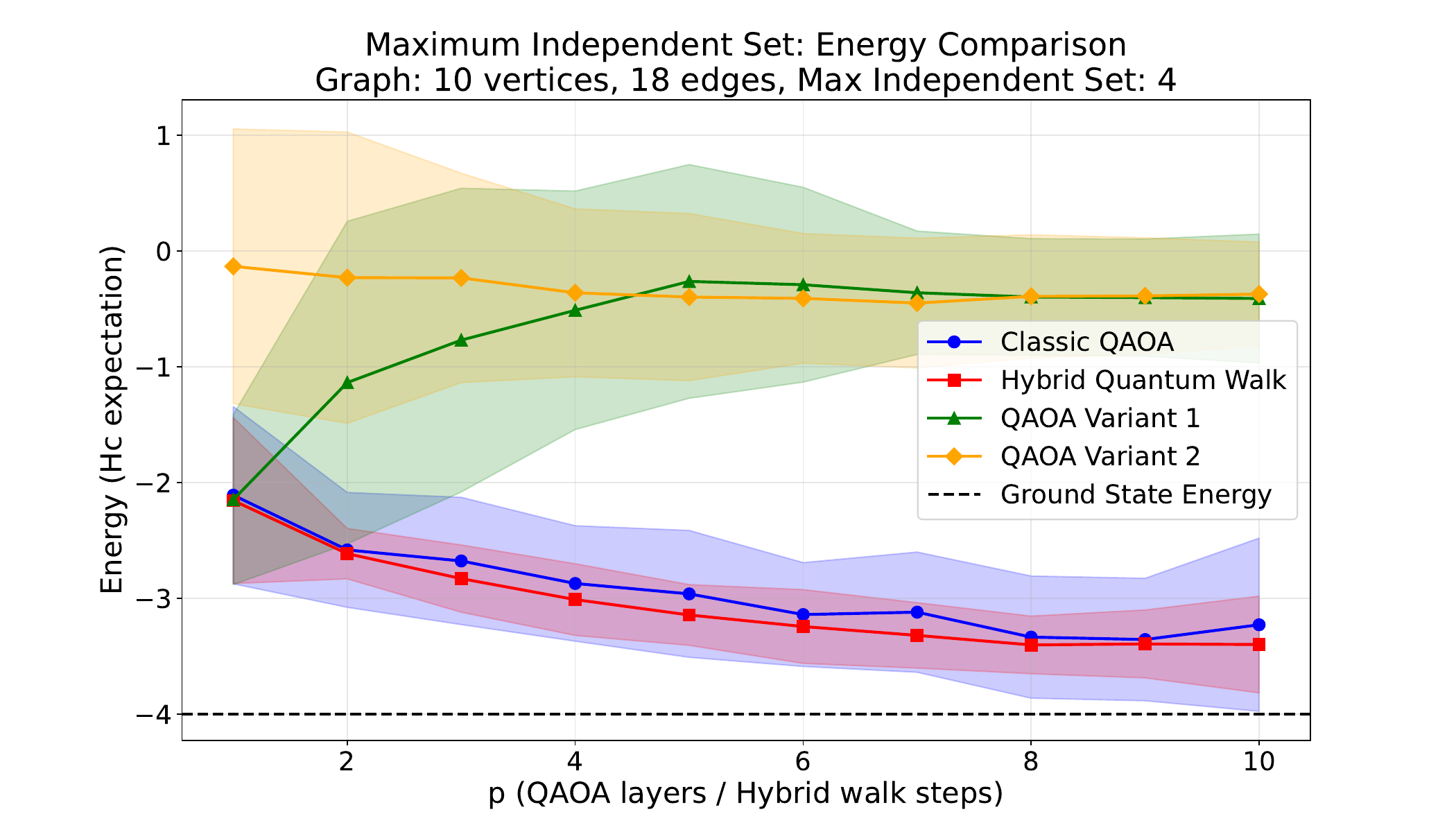}
\caption{Performance comparison on a 10-vertex MIS graph. The vertical axis shows the expectation value of $H_C$, with ground state energy $4$. HQW again converges faster and reaches lower final energy.}
\label{fig:s10}
\end{figure}

FIG.~\ref{fig:s9} shows the results for the 12-vertex Max-Cut graph, where the ground state energy is $-24$. HQW converges faster than QAOA and achieves lower energy at low depths, giving better sample efficiency. FIG.~\ref{fig:s10} presents analogous results for the 10-vertex MIS graph. Again, HQW outperforms QAOA, with faster convergence and lower final energy. In both cases, HQW maintains the smallest standard deviation across initializations, showing consistent robustness.

These results confirm that the algebraic advantage of HQW over QAOA persists for the system sizes tested. The witness $|\mathcal{N}_{\min}|$ governs its magnitude. Scaling to larger systems, and to problem encodings where $|\mathcal{N}_{\min}|$ does not decay with $n$, are natural next steps.

\subsection{Summary of Numerical Findings}

Across all tested instances, HQW consistently matches or exceeds QAOA performance. On 50 random Max-Cut graphs, HQW achieves a 98\% win rate in average-performance comparison (49/50 instances) and a 68\% win rate in best-performance comparison (34/50 instances), with no case of systematic underperformance. The advantage is not replicable by simply reordering QAOA parameters, confirming that the coin-controlled path-superposition mechanism is essential. Across the 90 Max-Cut instances analyzed in Sec.~\ref{DLA}, $|\mathcal{N}_{\min}| > 0$ guarantees HQW has strictly greater expressive capacity than QAOA, and this advantage is realized as a measurable $d_{\rm expl}$ gap in $97.8\%$ of cases and consistent performance gains across all benchmark instances.

These results indicate that activating the full Jordan-Lie algebraic structure of the problem Hamiltonians, rather than only the Lie-algebraic component, is a practical design principle for near-term quantum optimization.

\section{Conclusion}
\label{con.}
This paper reveals a Jordan-Lie algebraic structure beyond the standard DLA of variational quantum circuits. The DLA governs single-path ans\"atze such as QAOA and correctly captures their expressivity. We show that when a circuit architecture introduces control degrees of freedom that superpose multiple evolution paths, as in HQW model recently proposed, a richer algebraic layer emerges. Theorem~1 proves that the algebraic closure of HQW takes the form $\mathfrak{g}_{\rm H} = \mathfrak{su}(2)\otimes\mathcal{L}_{\rm Q} \oplus I_c\otimes\mathcal{K}_{\rm Q}$, strictly containing the DLA of any single-path ans\"atze. The additional directions originate from Jordan products $i\{H_c, H_b\}$ that lie outside the Lie closure and are algebraically inaccessible to single-path circuits regardless of depth. This is a structural expansion of accessible directions, not a matter of parameter count or layer depth.

The Jordan product negativity $|\mathcal{N}_{\min}|$ provides a classically computable witness for this enrichment, obtained by diagonalizing a single $N\times N$ matrix. It serves two purposes. First, $|\mathcal{N}_{\min}|>0$ guarantees that the Jordan-Lie algebra is strictly larger than the DLA. Second, its magnitude correlates with the strength of the algebraic advantage. Three independent measures, the DLA dimension ratio (Table~\ref{tab:dla_dims}), the fraction of $\mathfrak{su}(2^{n+1})$ occupied by $\mathfrak{g}_H$, and the effective exploration dimension $d_{\rm expl}$ (Fig.~\ref{fig:4}), all track the magnitude of $|\mathcal{N}_{\min}|$. For the standard Max-Cut encoding with an $X$-mixer, $|\mathcal{N}_{\min}|$ decays as a consequence of the two-local structure of the Hamiltonians, and the measured algebraic advantage moderates correspondingly. This decay is specific to the Hamiltonian pair, not a limitation of the Jordan-Lie framework. The witness captures both the existence and the effective range of Jordan-product enrichment, and its magnitude can be preserved at larger scales through alternative problem encodings or mixer design.

This consistent behavior across algebraic, geometric, and optimization-level measures has implications beyond HQW. First, it identifies a fundamental depth barrier for single-path ans\"atze. Jordan product directions remain inaccessible regardless of layer count, and HQW breaks this barrier in a single step. Second, both QAOA and HQW remain free of barren plateaus at the scales studied, since $\dim(\mathfrak{g}_H)$ stays polynomial in $n$, consistent with DLA theory~\cite{larocca2022dla,62} and results on overparametrization in quantum neural networks~\cite{55}. The larger algebra of HQW distributes gradients across more directions, producing a richer but still trainable optimization landscape. Third, the optimal coin operator from Pontryagin's minimum principle is generally not the Pauli-X gate, confirming from control theory that the path-superposition mechanism is essential.

Across 100+ problem instances, the algebraic predictions translate consistently into measured performance. HQW achieves a $1.60\times$ larger $d_{\rm expl}$ than QAOA in $97.8\%$ of 90 Max-Cut cases, independent of graph density. In $100+$ direct benchmarks, HQW outperforms QAOA on $98\%$ of medium-density graphs and all near-complete and  complete graphs. Component analysis confirms that the gain originates from coin-controlled path superposition and cannot be replicated by parameter reordering. The commuting limit provides the negative control. When $|\mathcal{N}_{\min}| = 0$, the $d_{\rm expl}$ gap collapses to within numerical noise and both ans\"atze reach the exact ground state, confirming that Jordan-product enrichment is the operative mechanism.

The Jordan-Lie framework is not specific to Max-Cut or to HQW. Any pair of non-commuting Hamiltonians carries a Jordan-algebraic structure, and any circuit architecture with auxiliary-qubit-controlled Hamiltonian evolutions can activate it. The $|\mathcal{N}_{\min}|$ witness can be computed classically before deploying quantum resources, identifying problem instances where activating the full Jordan-Lie structure is warranted. The algebraic advantage can be preserved at larger scales by designing problem encodings whose Jordan products retain non-negligible $|\mathcal{N}_{\min}|$, or by optimizing the mixer Hamiltonian to maximize $|\mathcal{N}_{\min}|$ for a given problem Hamiltonian. Both strategies are guided by the witness framework introduced here. Generalizing this Jordan-Lie design principle to other VQA architectures and problem classes is a direction for future work.

\section*{Data availability statement}
The data that support the findings of this study are openly available at \url {https://doi.org/10.5281/zenodo.18983535.}

\section*{Competing interests}
The authors declare that they have no known competing financial interests or personal relationships that could have appeared to influence the work reported in this paper.

\section*{References}

\bibliographystyle{unsrt} 
\bibliography{ref}

\section*{Appendix}

\appendix
\section{Pontryagin Minimum Principle: Full Derivation}
\label{app:pmp}

\noindent This appendix provides the complete derivation of the optimal coin operator via Pontryagin's minimum principle, supporting the compressed presentation in Sec.~\ref{Optimal coin}.

\begin{figure}[htbp]
  \centering
  \includegraphics[width=0.7\textwidth]{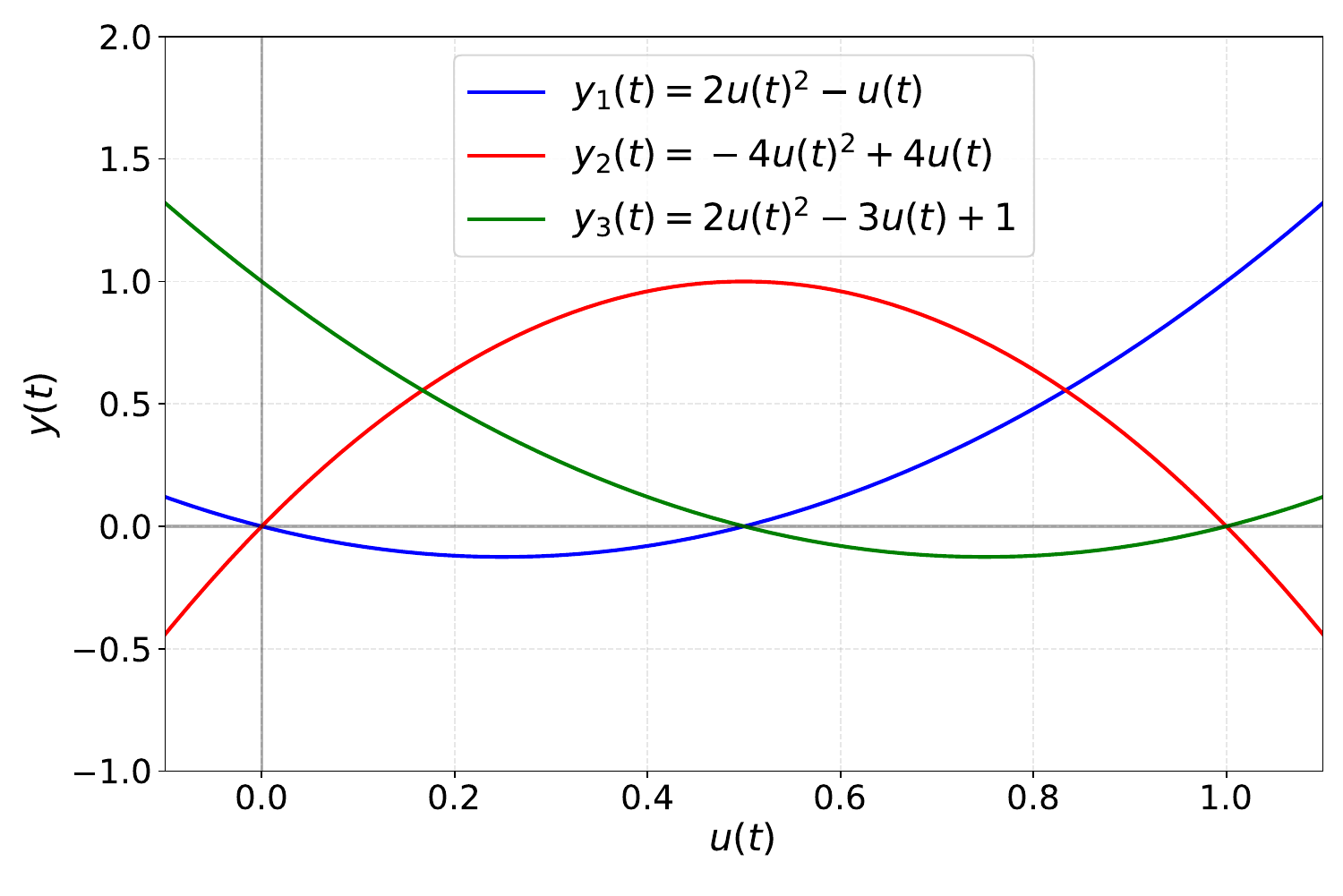}
  \caption{Function graphs of $y_1(t)$ (blue), $y_2(t)$ (red), and $y_3(t)$ (green) with respect to the control parameter $u(t)$.}
  \label{fig:s2}
\end{figure}

To apply PMP to the HQW dynamics, we design a time-dependent Hamiltonian that interpolates through three stages: coin rotation, evolution under $H_b$, and evolution under $H_c$, controlled by a parameter $u(t)\in[0,1]$:
\begin{align}
H(t)=y_{1}(t)|1\rangle\langle1|\otimes H_{c}+y_{2}(t)|0\rangle\langle0|\otimes H_{b} +y_{3}(t)(n_{x}(t)X+n_{y}(t)Y+n_{z}(t)Z)\otimes I,
\end{align}
where
\begin{align}
y_{1}(t)=2u(t)^{2}-u(t),\quad y_{2}(t)=-4u(t)^{2}+4u(t),\quad y_{3}(t)=2u(t)^{2}-3u(t)+1.
\end{align}
The function graphs of $y_{1}(t)$, $y_{2}(t)$, and $y_{3}(t)$ with respect to $u(t)$ are shown in FIG.~\ref{fig:s2}. When $u(t)$ changes slowly from $0$ to $1$, this describes an adiabatic evolution with an intermediate Hamiltonian. At $u(t)=0$, $y_{1}=y_{2}=0$ and $y_{3}=1$, giving the initial Hamiltonian $(n_{x}X+n_{y}Y+n_{z}Z)\otimes I$ (coin rotation). At $u(t)=1$, $y_{2}=y_{3}=0$ and $y_{1}=1$, giving the final Hamiltonian $|1\rangle\langle 1|\otimes H_{c}$. At $u(t)=\frac{1}{2}$, $y_{1}=y_{3}=0$ and $y_{2}=1$, giving the intermediate Hamiltonian $|0\rangle\langle 0|\otimes H_{b}$.

HQW is a discrete analogue of this adiabatic process. The control parameter $u(t)$ takes discrete values $\{0,\frac{1}{2},1\}$ corresponding to the three stages of each HQW step: coin operation, $H_b$ evolution, and $H_c$ evolution. The state $|\psi(t)\rangle$ evolves via the Schr\"odinger equation
\begin{equation}
|\dot{\psi}(t)\rangle=-iH(t)|\psi(t)\rangle.
\end{equation}

The three PMP prerequisites are:
\begin{enumerate}
    \item \textbf{Dynamical equation.} $|\dot\psi(t)\rangle=-iH(t)|\psi(t)\rangle$, $|\psi(t_0)\rangle=|\psi(0)\rangle$. The manifold $M$ is the unit sphere in $\mathbb{C}^n$.
    \item \textbf{Control input.} $\boldsymbol{v}(t)=(u(t),\boldsymbol{n}(t))\in\mathcal{U}=\{0, \frac{1}{2}, 1\}\times\{\boldsymbol{x}\in\mathbb{R}^3,\|\boldsymbol{x}\|_2=1\}\subset \mathbb{R}^4$.
    \item \textbf{Objective function.} $J=\langle\psi(t_f)|I\otimes H_c|\psi(t_f)\rangle$.
\end{enumerate}

The control Hamiltonian is
\begin{align}
\mathcal{H}=y_1(t)\Phi_{|1\rangle\langle1|\otimes H_c}(t)+y_2(t)\Phi_{|0\rangle\langle0|\otimes H_b}(t)
+y_3(t)\Phi_{(n_x X+n_y Y+n_z Z)\otimes I}(t),
\end{align}
where $\Phi_A(t) = i\langle k(t)|A|\psi(t)\rangle + \mathrm{c.c.}$, and the adjoint state $|k(t)\rangle$ satisfies
\begin{equation}
|\dot{k}(t)\rangle=-iH(t)|k(t)\rangle,\quad |k(t_{f})\rangle=I\otimes H_{c}|\psi(t_{f})\rangle.
\end{equation}

To maximize the objective $J$, we maximize the control Hamiltonian $\mathcal{H}$ subject to the unit-vector constraint $n_x^2+n_y^2+n_z^2=1$. Using the Lagrange multiplier method, define
\begin{equation}
L(n_x, n_y, n_z) = -\mathcal{H} + \mu(n_x^2 + n_y^2 + n_z^2 - 1).
\end{equation}
The stationarity conditions are:
\begin{align}
\frac{\partial L}{\partial n_{x}} &= -y_{3}(t)\Phi_{X\otimes I}(t) + 2\mu n_{x} = 0,\nonumber\\
\frac{\partial L}{\partial n_{y}} &= -y_{3}(t)\Phi_{Y\otimes I}(t) + 2\mu n_{y} = 0,\nonumber\\
\frac{\partial L}{\partial n_{z}} &= -y_{3}(t)\Phi_{Z\otimes I}(t) + 2\mu n_{z} = 0,
\end{align}
together with $n_{x}^{2}+n_{y}^{2}+n_{z}^{2}=1$. Solving for $n_x,n_y,n_z$ and the multiplier $\mu$ yields
\begin{equation}
\frac{1}{2\mu}=\frac{1}{|y_{3}(t)|\sqrt{\Phi_{X\otimes I}(t)^{2}+\Phi_{Y\otimes I}(t)^{2}+\Phi_{Z\otimes I}(t)^{2}}}.
\end{equation}

When the control parameter $u(t)=0$ (i.e., $y_{3}(t)=1$), the optimal coin parameters are
\begin{align}
n_{x}(t)&=\frac{\Phi_{X\otimes I}(t)}{\sqrt{\Phi_{X\otimes I}(t)^{2}+\Phi_{Y\otimes I}(t)^{2}+\Phi_{Z\otimes I}(t)^{2}}}, \\
n_{y}(t)&=\frac{\Phi_{Y\otimes I}(t)}{\sqrt{\Phi_{X\otimes I}(t)^{2}+\Phi_{Y\otimes I}(t)^{2}+\Phi_{Z\otimes I}(t)^{2}}}, \\
n_{z}(t)&=\frac{\Phi_{Z\otimes I}(t)}{\sqrt{\Phi_{X\otimes I}(t)^{2}+\Phi_{Y\otimes I}(t)^{2}+\Phi_{Z\otimes I}(t)^{2}}}.
\end{align}
The optimal coin rotation axis therefore aligns with the instantaneous sensitivity vector $(\Phi_X, \Phi_Y, \Phi_Z)$ in the coin space. In general this axis is not the Pauli-$X$ direction, confirming that QAOA's Pauli-$X$ coin does not saturate the optimum within the HQW family.

\section{Dynamical Lie Algebra Details}
\label{app:dla}

Since for any $2\times 2$ unitary operator $U$, there exists real numbers $\alpha$, $\beta$, $\gamma$, and $\delta$, such that $U=e^{i\alpha}e^{-i\beta Z/2}e^{-i\gamma Y/2}e^{-i\delta Z/2}$, DLA of HQW is $\mathfrak{g}_{\text{H}}=\langle iY\otimes I_{p},iZ\otimes I_{p},i|1\rangle\langle 1|\otimes H_{c},i|0\rangle\langle 0|\otimes H_{b}\rangle_{\text{Lie}}$, where $I_{p}$ is the identity operator of the position space.

To determine the algebraic structure of the DLA for HQW, we first present the definition of Jordan-Lie algebra.

\textbf{Definition 2 (Jordan-Lie algebra).} A Jordan-Lie algebra $\mathcal{L}$ is a real vector space equipped with two bilinear maps $\circ$ and $[\cdot,\cdot]:\mathcal{L}\times\mathcal{L}\rightarrow\mathcal{L}$, satisfying the following conditions:
\begin{enumerate}
\item $A\circ B=B\circ A$ and $[A,B]=-[B,A]$ for all $A,B\in\mathcal{L}$.
\item $[A,B\circ C]=[A,B]\circ C+B\circ[A,C]$ and $[A,[B,C]]=\left[[A,B],C\right]+[B,[A,C]]$ for all $A,B,C\in\mathcal{L}$.
\item $(A\circ B)\circ C-A\circ(B\circ C)=h^{2}[[A,C],B]$ for some $h^{2}\in\mathbb{R}$.
\end{enumerate}

According to the definition of the Jordan-Lie algebra, we next present a lemma to describe an algebra $\mathcal{L}_{\text{Q}}$.

\textbf{Lemma 1.} $\mathcal{L}_{\text{Q}}$ is the real vector space generated linearly by $\{iH_{c},iH_{b},iI_{p}\}$ equipped with two bilinear maps $i\{\cdot,\cdot\}$ and $[\cdot,\cdot]:\mathcal{L}_{\text{Q}}\times\mathcal{L}_{\text{Q}}\rightarrow\mathcal{L}_{\text{Q}}$, where $\{\cdot,\cdot\}$ and $[\cdot,\cdot]$ are the anti-commutator and the commutator, respectively, then $\mathcal{L}_{\text{Q}}$ is a Jordan-Lie algebra.

\textbf{Proof.} $\mathcal{L}_\text{Q}$ is the real vector space generated by $\{iH_c, iH_b, iI_p\}$.
    For any $A$, $B\in\mathcal{L}_\text{Q}$, $i\{A,B\}=iAB+iBA=i\{B,A\}$ and $[A,B]=AB-BA=-[B,A]$. For any $A$, $B$ and $C\in\mathcal{L}_\text{Q}$, 
    \begin{align}
        &[A,i\{B,C\}]=[A,iBC+iCB]=iABC+iACB-iBCA-iCBA.
    \end{align}
and
    \begin{align}
        i\{[A,B],C\}+i\{B,[A,C]\}&=i\{AB-BA,C\}+i\{B,AC-CA\}\nonumber\\
        &=iABC+iACB-iBCA-iCBA.       
    \end{align}
Thus, $[A,i\{B,C\}]=i\{[A,B],C\}+i\{B,[A,C]\}$. Since $[\cdot,\cdot]$ is the commutator, it is clear that $[A,[B,C]]=[[A,B],C]+[B,[A,C]]$. 
\begin{align}
        &i\{i\{A,B\},C\}-i\{A,i\{B,C\}\}\nonumber\\
        &=i\{iAB+iBA,C\}-i\{A,iBC+iCB\}\nonumber\\
        &=-ABC-BAC-CAB-CBA+ABC+ACB+BCA+CBA\nonumber\\
        &=-BAC-CAB+ACB+BCA=1\cdot[[A,C],B].     
    \end{align}
Therefore, $\mathcal{L}_\text{Q}$ is a Jordan-Lie algebra. \hfill $\square$

Finally, we present the specific form of the DLA of HQW in Theorem 1. 

\textbf{Theorem 1.} The dynamical Lie algebra of HQW is $\mathfrak{g}_{\text{H}}=\mathfrak{su}(2)\otimes\mathcal{L}_{\text{Q}}\oplus I_{c}\otimes\mathcal{K}_{\text{Q}}$, where $I_{c}$ is the identity operator of the coin space and $\mathcal{K}_{\text{Q}}=[\mathcal{L}_{\text{Q}},\mathcal{L}_{\text{Q}}]\oplus\operatorname{span}\{iH_{c},iH_{b}\}$.

\textbf{Proof.} Since $iY\otimes I_{p},iZ\otimes I_{p}\in\mathfrak{g}_{\text{H}}$, $-\frac{1}{2}[iY\otimes I_{p},iZ\otimes I_{p}]=iX\otimes I_{p}\in\mathfrak{g}_{\text{H}}$. $-[iX\otimes I_{p},i|1\rangle\langle 1|\otimes H_{c}]=iY\otimes H_{c}\in\mathfrak{g}_{\text{H}}$, so $-\frac{1}{2}[iX\otimes I_{p},iY\otimes H_{c}]=iZ\otimes H_{c}$ and $-\frac{1}{2}[iY\otimes H_{c},iZ\otimes I_{p}]=iX\otimes H_{c}\in\mathfrak{g}_{\text{H}}$. And we can get $iZ\otimes H_{c}+2i|1\rangle\langle 1|\otimes H_{c}=iI_{c}\otimes H_{c}\in\mathfrak{g}_{\text{H}}$. Therefore, $P\otimes iH_{c}\in\mathfrak{g}_{\text{H}}$, $P\in\{I_{c},X,Y,Z\}$. Similarly, $P\otimes iH_{b}\in\mathfrak{g}_{\text{H}}$, $P\in\{I_{c},X,Y,Z\}$. Thus, $\mathfrak{g}_{\text{H}}=\langle\mathcal{A}_{1}\rangle_{\text{Lie}}$, where $\mathcal{A}_{1}=\{P^{\times}\otimes iI_{p},P\otimes iH_{c},P\otimes iH_{b}:P^{\times}\in\{X,Y,Z\},P\in\{I_{c},X,Y,Z\}\}\rangle_{\text{Lie}}$. Since the commutators of $P^{\times}\otimes iI_{p}$ with other operators are $0$ or in $\mathcal{A}_{1}$, we only consider the operator commutator set of the set $\{P\otimes iH_{c},P\otimes iH_{b}:P\in\{I_{c},X,Y,Z\}\}$. Now we consider $[A_{1}\otimes B_{1},A_{2}\otimes B_{2}]=A_{1}A_{2}\otimes B_{1}B_{2}-A_{2}A_{1}\otimes B_{2}B_{1}$ for any $A_{1}$ and $A_{2}\in\{I_{c},X,Y,Z\}\}$ and two operators $B_{1}$ and $B_{2}$. There are $3$ cases as follows:

\begin{enumerate}
\item[(i)] If $A_{1}\neq A_{2}$ and $[A_{1},A_{2}]\neq 0$, such as $X$ and $Y$, then $\exists P^{\times}\in\{X,Y,Z\}\}$ such that $\frac{1}{2}[A_{1}\otimes B_{1},A_{2}\otimes B_{2}]=P^{\times}\otimes i\{B_{1},B_{2}\}$ or $-P^{\times}\otimes i\{B_{1},B_{2}\}$.
\item[(ii)] If $A_{1}\neq A_{2}$ and $[A_{1},A_{2}]=0$, then $A_{1}=I_{c}$ or $A_{2}=I_{c}$ and $\exists P^{\times}\in\{X,Y,Z\}\}$ such that $[A_{1}\otimes B_{1},A_{2}\otimes B_{2}]=P^{\times}\otimes[B_{1},B_{2}]$.
\item[(iii)] If $A_{1}=A_{2}$, then $[A_{1}\otimes B_{1},A_{2}\otimes B_{2}]=I_{c}\otimes[B_{1},B_{2}]$.
\end{enumerate}

The set generated by $iH_{c}$ and $iH_{b}$ through operations $[\cdot,\cdot]$ and $i\{\cdot,\cdot\}$ is denoted as $\mathcal{L}^{\prime}_{\text{Q}}$. According to the cases (i) and (ii), we can get $\mathfrak{g}_{\text{H}}=\langle\{P^{\times}\otimes iI_{p},P^{\times}\otimes B,I_{c}\otimes iH_{c},I_{c}\otimes iH_{b}:P^{\times}\in\{X,Y,Z\},B\in\mathcal{L}^{\prime}_{\text{Q}}\}\rangle_{\text{Lie}}$. Since $i\{iI_{p},B\}=-2B$ and $[iI_{p},B]=0$, the real vector space generated by $\mathcal{L}^{\prime}_{\text{Q}}\cup\{iI_{p}\}$ is $\mathcal{L}_{\text{Q}}$. Now we can obtain $\mathfrak{g}_{\text{H}}=\langle\{P^{\times}\otimes B,I_{c}\otimes iH_{c},I_{c}\otimes iH_{b}:P^{\times}\in\{X,Y,Z\},B\in\mathcal{L}_{\text{Q}}\}\rangle_{\text{Lie}}$. According to the case (iii), we can get $\mathfrak{g}_{\text{H}}=\langle\mathcal{A}_{2}\rangle_{\text{Lie}}$, where $\mathcal{A}_{2}=\{P^{\times}\otimes B,I_{c}\otimes B^{\prime}:P^{\times}\in\{X,Y,Z\},B\in\mathcal{L}_{\text{Q}},B^{\prime}\in\{[A,B]:A,B\in\mathcal{L}_{\text{Q}}\}\cup\{iH_{c},iH_{b}\}\}$. It is clear that the commutator of any two elements of $\mathcal{A}_{2}$ is still in $\mathcal{A}_{2}$. And the real vector space generated by $\mathcal{A}_{2}$ is $\mathfrak{su}(2)\otimes\mathcal{L}_{\text{Q}}\oplus I_{c}\otimes\mathcal{K}_{\text{Q}}$. Therefore, $\mathfrak{g}_{\text{H}}=\mathfrak{su}(2)\otimes\mathcal{L}_{\text{Q}}\oplus I_{c}\otimes\mathcal{K}_{\text{Q}}$. \hfill $\square$

\section{Exploration Strength: Computational Details and Further Analysis}
\label{app:exploration}

\noindent This appendix provides the computational details of the $d_{\rm
expl}$ metric, extended numerical breakdowns, and the full discussion of
implications referred to in Sec.~\ref{DLA}.

\subsection{Computational Method}

For each variational parameter $\theta_i$ at the optimized point
$\boldsymbol{\theta}^*$, we perturb by $+\varepsilon$ ($\varepsilon =
10^{-4}$) and compute the state $|\psi(\theta_i + \varepsilon)\rangle$.
The perturbed state is phase-aligned to the reference state
$|\psi(\boldsymbol{\theta}^*)\rangle$ via
$\langle\psi(\boldsymbol{\theta}^*)|\psi(\theta_i+\varepsilon)\rangle
\to |\langle\cdots\rangle|$ to remove the irrelevant global phase. The
one-sided finite-difference tangent vector is
$\boldsymbol{t}_i = (|\psi(\theta_i+\varepsilon)\rangle -
|\psi(\boldsymbol{\theta}^*)\rangle)/\varepsilon$. The matrix
$Q = [\boldsymbol{t}_1, \ldots, \boldsymbol{t}_m]$ with $m = 2p_{\rm
QAOA}$ for QAOA and $m = 3p_{\rm HQW}$ for HQW is formed, its
singular-value decomposition $Q = U \Sigma V^\dagger$ is computed, and
the effective dimension is evaluated as the participation ratio
$d_{\rm expl} = (\sum_i \sigma_i)^2 / \sum_i \sigma_i^2$
(Eq.~\eqref{eq:eff_rank}). This continuous measure is bounded between
$1$ and $m$ and captures the soft dimensionality of the tangent space:
if all singular values are comparable, $d_{\rm expl} \approx m$. If one
direction dominates, $d_{\rm expl} \approx 1$.

\subsection{Size Scaling and Graph Structure Effects}

The $d_{\rm expl}$ gap varies with system size: it is largest at $n=7$
(mean $\Delta d_{\rm expl} = 1.69$, ratio $1.74\times$), decreasing to
$1.25$ ($1.53\times$) at $n=8$ and $1.16$ ($1.52\times$) at $n=9$. This
reflects the growth of the ambient Hilbert space ($D = 2^n$): the algebraic
enrichment from the Jordan product direction provides a stable contribution,
which becomes a smaller fraction of the total dimensions as $n$ increases.
When normalized by the Hilbert space dimension: $S = d_{\rm expl}/D$ with $D_{\rm QAOA}=2^{n}$, $D_{\rm HQW}=2^{n+1}$,
$\Delta S = S_{\rm H} - S_{\rm Q}$ is predominantly negative due to the
factor-of-2 larger denominator for HQW, a normalization artifact that
masks the consistent raw $d_{\rm expl}$ advantage. We therefore report both
raw $d_{\rm expl}$ and normalized $S$ values for transparency.

The gap also varies with density class: the mean $\Delta d_{\rm expl}$ is
$1.19 \pm 0.48$ for sparse graphs (density $< 0.3$), $1.44 \pm 0.61$ for
medium-density graphs, and $1.59 \pm 0.80$ for near-complete graphs
(density $> 0.9$). Although the trend is weakly positive, the overall
density--gap correlation is negligible (Pearson $r = 0.058$, $p = 0.59$,
Spearman $\rho = 0.036$, $p = 0.74$), so density can be excluded as a
driving factor for the observed advantage.  Regarding graph structure, near-complete graphs ($K_n - 1$,
$K_n - 2$) produce larger average gaps than perfect complete graphs $K_n$
at $n = 7, 8$, hinting that breaking the full permutation symmetry of
$K_n$ may expose additional accessible directions in the tangent space.

\subsection{The Negative Side of the Witness}

All 90 benchmarked Max-Cut graphs satisfy $|\mathcal{N}_{\min}| > 0$. This is not accidental: for QAOA-type Hamiltonians where $H_b = \sum_i X_i$ has zero diagonal in the basis diagonalizing $H_c$, one has $\operatorname{Tr}(H_c \circ H_b) = \operatorname{Tr}(H_c H_b) = 0$. The eigenvalues of $H_c \circ H_b$ therefore sum to zero, forcing both positive and negative eigenvalues or all zero, so the minimum eigenvalue satisfies $\mathcal{N}_{\min} \in [-1, 0]$. For any non-trivial connected graph, $H_c$ and $H_b$ do not commute, and $|\mathcal{N}_{\min}| > 0$ follows. The witness is thus always “on” for the problem class studied.

To validate the negative side of the witness, we take two complementary approaches. First, we construct the commuting limit: $H_c = H_b = \sum_i X_i$. For even $n$, the matrix $\sum_i X_i$ has a zero eigenvalue, so the normalized Jordan product $H_c \circ H_b = H_c^2$ has minimum eigenvalue exactly zero, giving $|\mathcal{N}_{\min}| = 0$. For odd $n$, $|\mathcal{N}_{\min}| > 0$ since the smallest absolute eigenvalue of $\sum_i X_i$ is 1, yet commutation prevents any Jordan-product enrichment of the DLA. We test $n = 5, 6, 7$ with $p_{\rm QAOA}=2$, $p_{\rm HQW}=4$, 5 random restarts each with 100 Adam steps. Across all system sizes, $d_{\rm expl}^{\rm Q} = 1.0000$ exactly and $d_{\rm expl}^{\rm H} = 1.0022$ ($n=5$), $1.0027$ ($n=6$), $1.0031$ ($n=7$), each within $0.003$ of unity. Both ans\"atze reach the exact ground-state energy $-n$ in every trial. The $\Delta d_{\rm expl}$ gap is therefore zero to within numerical noise, regardless of whether $|\mathcal{N}_{\min}|$ is strictly zero or positive. Commutation alone suffices to collapse the advantage, and the witness correctly signals the absence of Jordan-product enrichment whenever $|\mathcal{N}_{\min}| = 0$.

Second, we examine the $n$-dependence of $|\mathcal{N}_{\min}|$. As the system size grows, the eigenvalues of the normalized Jordan product are diluted across an exponentially larger Hilbert space: $|\mathcal{N}_{\min}|$ decreases from $\sim\!0.042$ at $n=7$ to $\sim\!0.023$ at $n=8$, $\sim\!0.015$ at $n=9$, and $\sim\!0.006$ at $n=10$, roughly halving with each added qubit. The $d_{\rm expl}$ ratio narrows correspondingly from $1.74\times$ ($n=7$) to $1.53\times$ ($n=8$) to $1.52\times$ ($n=9$). Within each fixed $n$, $|\mathcal{N}_{\min}|$ varies minimally across graph structures, all giving values within $\sim\!15\%$ of each other, so per-$n$ correlation between $|\mathcal{N}_{\min}|$ and $\Delta d_{\rm expl}$ is negligible. This is consistent with the witness being binary rather than quantitative: $|\mathcal{N}_{\min}| > 0$ guarantees the existence of an advantage but does not predict its magnitude.

The two tests together, the commuting limit where $|\mathcal{N}_{\min}| = 0$ yields no advantage and the $n$-scaling where $|\mathcal{N}_{\min}|$ and $d_{\rm expl}$ gap shrink in parallel, provide a closed-loop validation of the witness. We emphasize that this $n$-decay of $|\mathcal{N}_{\min}|$ is a property of the standard Max-Cut encoding with an $X$-mixer, caused by the two-local structure of the Hamiltonians. It is not intrinsic to the Jordan-Lie framework. For problem Hamiltonians with multi-body interactions, or for alternative encodings of the same combinatorial problem, $|\mathcal{N}_{\min}|$ need not decay with $n$. \ref{app:nmin_scaling} provides numerical evidence supporting this encoding-specific interpretation.

\subsection{Encoding Dependence of $|\mathcal{N}_{\min}|$: Numerical Evidence}
\label{app:nmin_scaling}

To verify that the decay of $|\mathcal{N}_{\min}|$ is a property of the specific Hamiltonian pair rather than an inherent limitation of the Jordan-Lie framework, we compute $|\mathcal{N}_{\min}|$ for $n = 3$ to $10$ across several Hamiltonian configurations.

\textbf{Scenario A (standard):} Path-graph Max-Cut $H_c = \sum_{i=1}^{n-1}(I - Z_i Z_{i+1})/2$ with the standard $X$-mixer $H_b = \sum_i X_i$. This is the configuration used in Table~1. $|\mathcal{N}_{\min}|$ decays from $0.184$ at $n=3$ to $0.0034$ at $n=10$, roughly halving at each step with a decay ratio of approximately $0.53$ to $0.63$ per added qubit. The decay is caused by the sparse structure of the Jordan product: each $Z_i Z_{i+1}$ term anticommutes with $X_i$ and $X_{i+1}$ and commutes with the remaining $n-2$ $X$ operators, producing $\mathcal{O}(n^2)$ non-zero Pauli strings in a $2^n \times 2^n$ matrix.

\textbf{Scenario B (denser $H_c$):} Complete-graph Max-Cut with the same $X$-mixer. At $n=3$, $|\mathcal{N}_{\min}|$ is $1.05\times$ that of Scenario A, but the ratio decreases monotonically with $n$, crossing below $1.0$ at $n=6$ ($0.98$) and reaching $0.93$ at $n=10$. The complete graph adds more $ZZ$ terms, but the Frobenius normalization absorbs the increased norm, so the per-edge eigenvalue contribution decreases. The decay rate per qubit ($0.53$--$0.62\times$) is indistinguishable from Scenario A, since the locality structure, 2-local $ZZ$ plus 1-local $X$, is unchanged; only the absolute prefactor differs slightly.

\textbf{Scenario C (modified mixer):} Path-graph Max-Cut with a mixer containing a small identity component, $H_b = (1-\alpha)\sum_i X_i + \alpha I$ ($\alpha = 0.3$). The identity contribution breaks the zero-trace constraint $\operatorname{Tr}(H_c \circ H_b) = 0$, with a trace of approximately $0.35$ for all $n$. However, $|\mathcal{N}_{\min}|$ is $0.67\times$--$0.78\times$ that of Scenario A, i.e., 22--33\% smaller at each $n$. Interestingly, the ratio C/A increases with $n$ from $0.67$ at $n=3$ to $0.78$ at $n=10$, indicating that $|\mathcal{N}_{\min}|$ decays more slowly when the trace is non-zero. This demonstrates that both the magnitude and the scaling of $|\mathcal{N}_{\min}|$ are sensitive to the mixer structure, not solely to $H_c$.

Two distinct effects emerge. Scenario B shifts the absolute value of $|\mathcal{N}_{\min}|$ without altering the per-qubit decay rate: the locality structure of the Jordan product is unchanged, so the eigenvalue dilution with $n$ follows the same $0.53$--$0.63\times$ scaling. Scenario C changes both the magnitude and the decay rate: the ratio C/A increases monotonically with $n$, proving that the scaling exponent itself is mixer-dependent. The decay observed for the standard Max-Cut $+$ $X$-mixer is therefore a property of the specific two-local, off-diagonal Hamiltonian pair, not an intrinsic feature of Jordan products or of the Jordan-Lie framework. Preserving $|\mathcal{N}_{\min}|$ at larger system sizes is an encoding and mixer-design question, not a fundamental barrier.

\section{Additional Numerical Experiment Details}
\label{app:num_exp}

This appendix provides supplementary details that support the main findings presented in Sec.~\ref{num.exp.}.

\subsection{Problem Hamiltonians}

For a graph $G=(V,E)$, the Max-Cut problem is encoded using the standard Hamiltonian
\begin{equation}
H_c = \sum_{(i,j)\in E} \frac{I - Z_i Z_j}{2},
\end{equation}
with the mixer Hamiltonian $H_b = \sum_i X_i$.

The Maximum Independent Set problem is encoded as
\begin{equation}
H_c = -\sum_{i} n_i + \lambda \sum_{(i,j) \in E} n_i n_j,
\label{eq:hamiltonian_general}
\end{equation}
where $n_i = \frac{1}{2}(I - Z_i)$ represents the occupation number operator for vertex $i$, indicating whether vertex $i$ is included ($n_i=1$) or excluded ($n_i=0$) from the independent set. The parameter $\lambda > 0$ is a penalty coefficient that enforces the independence constraint, which is set to $\lambda = 1$ in our implementation. 

For $\boldsymbol{x}\in\mathbb{F}_{2}^{n}$, we have
\begin{equation}
H_{c}|\boldsymbol{x}\rangle=N(\boldsymbol{x})|\boldsymbol{x}\rangle,\quad H_{b}|\boldsymbol{x}\rangle=\sum_{\begin{subarray}{c}\boldsymbol{y}\in\mathbb{F}_{2}^{n}\\ |\boldsymbol{y}-\boldsymbol{x}|=1\end{subarray}}|\boldsymbol{y}\rangle,
\end{equation}
where $N(\boldsymbol{x})$ is the solution value corresponding to $\boldsymbol{x}$ and $|\boldsymbol{y}-\boldsymbol{x}|$ denotes the Hamming distance between the binary strings $\boldsymbol{x}$ and $\boldsymbol{y}$. Thus, we can use the $n$-dimensional hypercube $Q_{n}$ with a self-loop of weight $N(\boldsymbol{x})$ at vertex $\boldsymbol{x}$, with all self-loops labeled 1 and all other edges labeled 0, as the graph for the HQW. This clear graph structure also provides a solid theoretical foundation for future experimental implementation of HQW to solve this problem on physical platforms.
\subsection{Circuit Implementations}
The classic QAOA circuit uses only $n$ system qubits. After Hadamard initialization, each layer $l$ applies:
\begin{equation}
U(\gamma_{l},\beta_{l})=\exp(-i\gamma_{l}H_{c})\exp(-i\beta_{l}H_{b})
\end{equation}
where $\gamma_{l}$ and $\beta_{l}$ are variational parameters optimized classically. 

The hybrid quantum walk circuit employs $n+1$ qubits: $n$ system qubits representing graph vertices and $1$ coin qubit controlling the evolution. The circuit is initialized with Hadamard gates applied to all system qubits. For each step $s$, the following operations are performed:
\begin{enumerate}
\item Apply a $U_{3}(\theta_{s,2},\theta_{s,3},\theta_{s,4})$ gate on the coin qubit, where
\begin{equation}
U_{3}(\theta,\phi,\delta)=\begin{bmatrix}\cos(\theta/2)&-\exp(i\delta)\sin(\theta/2)\\ \exp(i\phi)\sin(\theta/2)&\exp(i(\phi+\delta))\cos(\theta/2)\end{bmatrix}.
\end{equation}
\item If the coin qubit is in state $|1\rangle$, apply $\exp(-i\theta_{s,0}H_{c})$ to system qubits. If the coin qubit is in state $|0\rangle$, apply $\exp(-i\theta_{s,1}H_{b})$ to system qubits.
\end{enumerate}
The controlled evolution is implemented using PennyLane's ctrl operation with conditional application of ApproxTimeEvolution. Each layer of QAOA adds two evolution processes corresponding to the two Hamiltonians respectively, while each step of HQW can only add one evolution process for either of the two Hamiltonians in each evolution path. Therefore, for the sake of fairness, we set the number of layers of QAOA as $p$ and the number of steps of HQW as $2p$.

We note a subtlety in the resource comparison. While the number of Hamiltonian-evolution gates is matched, each HQW step additionally applies a coin rotation $C_k$ and operates on an enlarged Hilbert space $\mathcal{H}_c \otimes \mathcal{H}_p$ of dimension $2^{n+1}$, compared to $2^n$ for QAOA. With $p$ QAOA layers and $2p$ HQW steps, the variational parameter counts are $m_{\rm QAOA} = 2p$ and $m_{\rm HQW} = 3 \cdot (2p) = 6p$. We chose this comparison because it equalizes the number of evolution operators, the dominant computational cost in near-term implementations, while acknowledging that HQW uses a moderately larger resource budget. Perfectly equalizing parameter count, circuit depth, and Hilbert-space dimension simultaneously is structurally impossible for ans\"atze with different Lie-algebraic generators. This asymmetry is common across the VQA comparison literature. Therefore, after obtaining the optimal time evolution parameters for the $2p$-step HQW through parameter optimization, the parameters of the two QAOA variants are selected in an interleaved manner to determine their $p$-layer quantum circuits. We employ the Adam optimizer \cite{53} with learning rate $0.1$. In summary, we benchmarked four algorithms:
\begin{enumerate}
\item \textbf{Standard QAOA}: The standard QAOA ansatz $U_b(\beta_p)U_c(\gamma_p)\cdots U_b(\beta_1)U_c(\gamma_1)$.
\item Hybrid Quantum Walk (HQW): The HQW operator 
\begin{align}
W(\gamma_k, \beta_k) = e^{-i|1\rangle\langle 1|\otimes H_c \gamma_k} e^{-i|0\rangle\langle 0|\otimes H_b \beta_k} (C_k \otimes I),\nonumber    
\end{align}
with $p$ steps.
\item QAOA Variant 1: A QAOA circuit where parameters $\gamma_k$ (for $H_c$) are taken from the odd-step HQW parameters and $\beta_k$ (for $H_b$) from the even-step HQW parameters.
\item QAOA Variant 2: A QAOA circuit with $\gamma_k$ from even-step and $\beta_k$ from odd-step HQW parameters.
\end{enumerate}
For HQW, the final expectation value is $\langle -I \otimes H_c \rangle$ for the Max-Cut problem and $\langle I \otimes H_c \rangle$ for the Maximum Independent Set problem. All circuits were optimized to minimize this expectation.

\subsection{Numerical Differentiation and Optimization Details}

All gradient computations use one-sided finite differences. For the Adam optimizer, the finite-difference step is $h = 10^{-5}$. For the tangent vectors in the $d_{\rm expl}$ computation (Eq.~\eqref{eq:eff_rank}), the finite-difference step is $\varepsilon = 10^{-4}$, with the perturbed state phase-aligned to the reference state to remove the irrelevant global phase before the difference is taken. The tangent vectors are assembled into the matrix $Q$, whose singular values $\sigma_i$ are obtained via SVD, and the effective rank is computed as the participation ratio $d_{\rm expl} = (\sum_i \sigma_i)^2 / \sum_i \sigma_i^2$.

Optimization uses the Adam optimizer~\cite{53} with learning rate $0.1$, $\beta_1 = 0.9$, $\beta_2 = 0.999$, running for $n_{\rm steps} = 150$ iterations per restart. For the 90-graph exploration analysis (Sec.~\ref{DLA}), $n_{\rm restarts} = 5$ random initializations are used, with the best cost across restarts selected. For the performance benchmarks (Sec.~\ref{num.exp.}), $n_{\rm restarts} = 100$ or 200 for the larger 12-vertex graphs and $n_{\rm steps} = 300$. All variational parameters are initialized uniformly from $[0, 2\pi]$ at each restart. Convergence is assessed by the final cost value across restarts. Across the 90-graph benchmark, the standard deviation of the final cost across the 5 restarts is typically below $10^{-3}$ of the mean, indicating stable convergence.

\subsection{Detailed Discussion of best-performance Improvements on 50 Graphs}

Here we provide additional details for the best-performance comparison on 50 Graphs. As noted in the main text, the experimental protocol is identical to that described for the best-performance comparison: QAOA depth $p=2$, HQW steps $2p=4$, 100 random initializations, 300 Adam steps with learning rate $0.1$, and we take the best $1-r$ over the initializations for each graph.
\begin{figure}[htbp]
\centering
\includegraphics[width=0.6\textwidth]{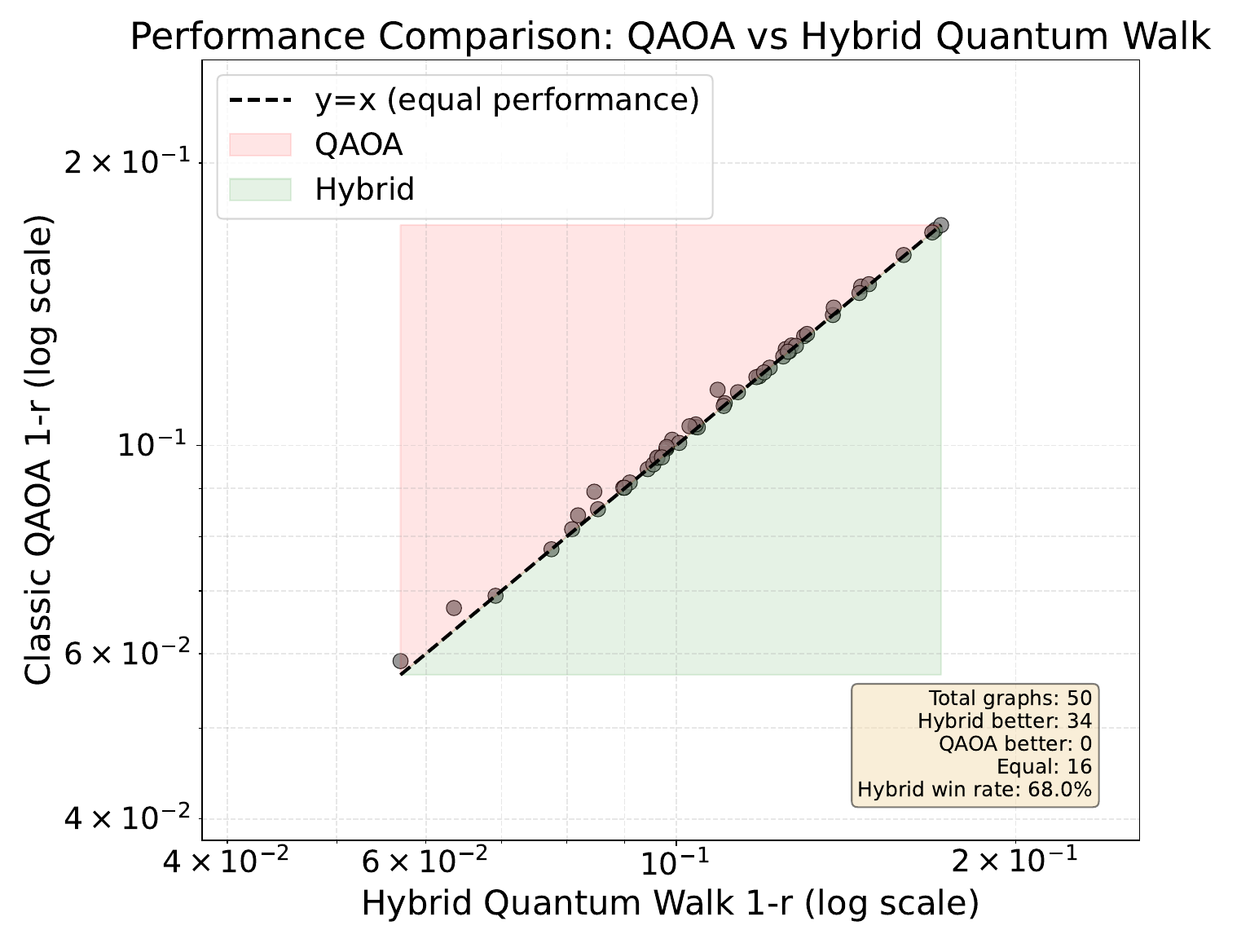}
\caption{A scatter plot of best $1-r$ (over 100 initializations) for QAOA vs. HQW on 50 Max-Cut graphs. All points lie on or above the diagonal under a tolerance of $10^{-6}$. Both axes show $1-r$ on a logarithmic scale.}
\label{fig:s3}
\end{figure}

\begin{figure}[htbp]
\centering
\includegraphics[width=0.6\textwidth]{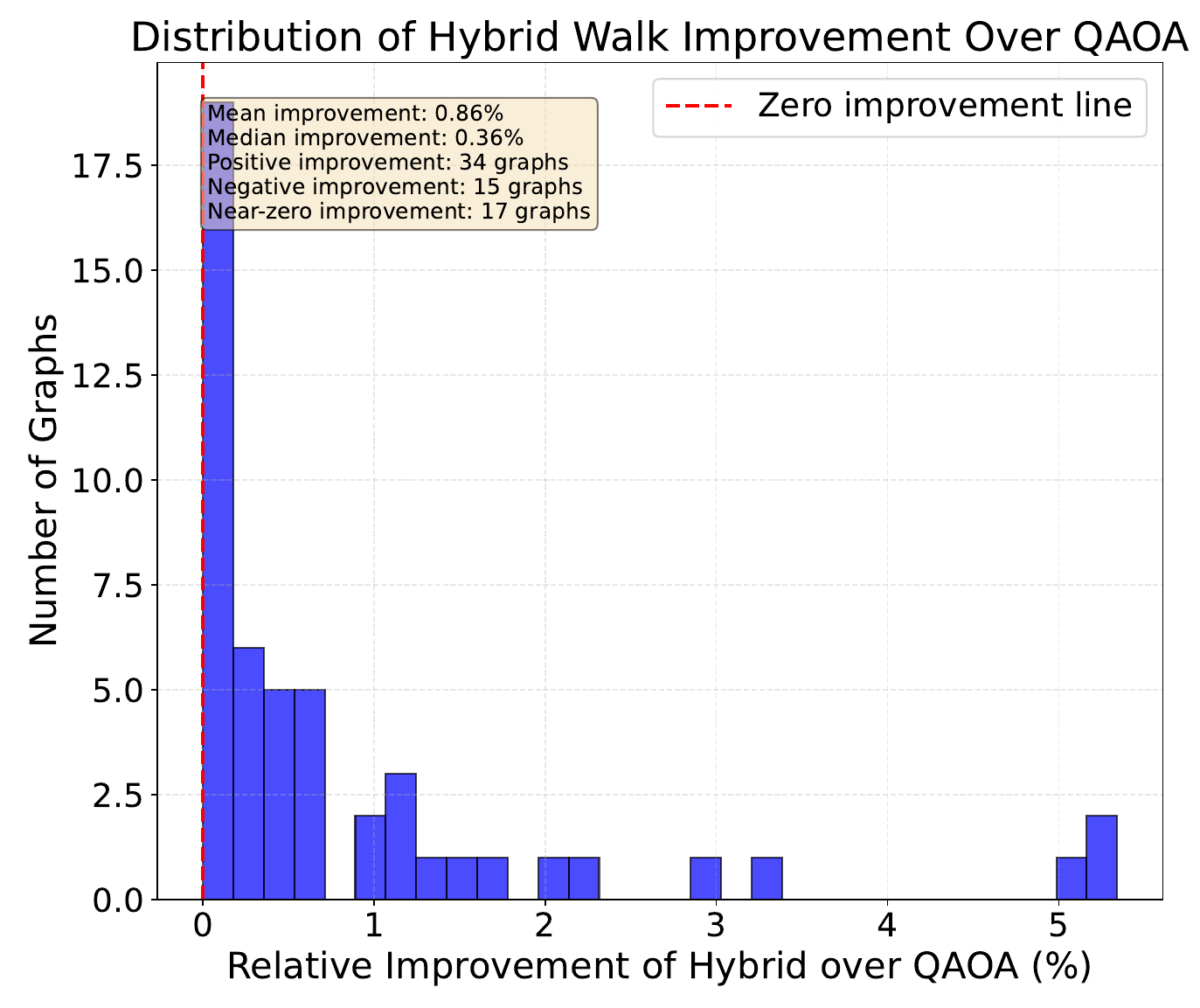}
\caption{Distribution of HQW's relative improvement over QAOA for the best-performance comparison. Mean: 0.86\%, median: 0.36\%, 34 positive, 15 negative (within $[-10^{-6},0)$), 17 near-zero improvements.}
\label{fig:s4}
\end{figure}
A scatter plot comparing the best $1-r$ values for each graph is shown in FIG.~\ref{fig:s3}. All points lie on or above the $y=x$ line, confirming that HQW never underperforms QAOA. 15 graphs showed small negative relative improvements in the histogram analysis (Fig.~\ref{fig:s4}). This occurs because the relative improvement is defined as $\frac{(1-r)_{\text{QAOA}}-(1-r)_{\text{HQW}}}{(1-r)_{\text{QAOA}}}\times 100\%$. When the absolute differences between the two algorithms are extremely small (less than $10^{-6}$), tiny numerical fluctuations can yield small negative values. These instances are still classified as “equal performance” under our tolerance criterion, as shown in the scatter plot (Fig.~\ref{fig:s3}) where all points lie on or above the diagonal within the tolerance. The distribution shows 17 graphs with near-zero improvements ($-0.1\%$ to $0.1\%$), comprising 15 graphs with negative improvements ($[-10^{-4}\%,0\%)$), 1 graph with $0\%$ improvement and 1 graph with positive improvement ($(0\%,0.1\%]$). 30 graphs show small improvements ($0.1\%$–$5\%$), and 3 graphs show medium improvements ($5\%$–$20\%$). No graphs exhibited improvements exceeding $20\%$. This distribution, together with the standard deviation of $1.34\%$, indicates that the advantage of HQW is consistent but modest in absolute magnitude for these small instances.

\end{document}